\documentclass[12pt]{article}
\let\ps=\psi

\def\\{\hfill\break} \let\==\equiv

\let\0=\noindent
\def\bra#1{{\langle#1|}}\def\ket#1{{|#1\rangle}}

\def\nn{\nonumber}

\def\qed{\hfill\raise1pt\hbox{\vrule height5pt width5pt depth0pt}}

\def\be{\begin{equation}}
\def\ee{\end{equation}}
\def\bea{\begin{eqnarray}}\def\eea{\end{eqnarray}}
\def\be{\begin{equation}}
\def\ee{\end{equation}}
\def\bea{\begin{eqnarray}}
\def\eea{\end{eqnarray}}

\def\hsp5{\hspace{5mm}}

\def\case#1/#2{\textstyle\frac{#1}{#2}}

\title{On the limits of quantum theory: \\
Contextuality and the quantum-classical cut}
\author{George F R Ellis\\
Mathematics Department, University of Cape Town.}

\begin{document}
\maketitle \vspace{0.1in}
\begin{abstract}
This paper is based on four assumptions: 1. Physical reality is made of linearly behaving components
  combined in non-linear ways. 2. Higher level behaviour emerges from this lower level structure.
  3. The way the lower level elements behaves depends on the context in which they are imbedded.
  4.  Quantum theory applies to the lower level entities.
An implication is that higher level effective laws, based in the
outcomes of non-linear combinations of lower level linear
interactions, will generically not be unitary; hence the
applicability of quantum theory at higher levels is strictly
limited. This leads to the view that both state vector preparation
and the quantum measurement process are crucially based in top-down
causal effects, and helps provide criteria for the Heisenberg cut
that challenge some views on Schr\"{o}dinger's cat.
\end{abstract}
\tableofcontents

\section{Quantum theory and classicality}
The classical to quantum relation is a key issue in understanding
how quantum theory applies to the real world. In order to make
progress in understanding this relation, it may well be profitable
to consider firstly the way complexity emerges from the underlying
physical relations, and secondly the way the operation of underlying
physical processes is contextually determined. This paper will make
the case that examining these issues of emergence and contextuality
helps clarify the nature of the classical-quantum cut, also known as
\emph{Heisenberg's cut} (\cite{WisMil10}:15), and hence the way that
non-quantum macro behaviour can emerge from underlying quantum
systems.\\

The basic viewpoint taken here is that physical theory must explain
not only what happens in carefully controlled laboratory
experiments, but also the commonplace features of life around us,
for which we have a huge amount of evidence in our daily lives. We
will set out this viewpoint in more detail in Section
\ref{sec:foundations} below, after first setting out the fundamental
quantum dilemma. Further sections will explore the ways that quantum
behaviour might emerge at higher levels of the hierarchy of
complexity; will suggest contexts where this will almost certainly
not be possible; and will explore the way contextual effects may
help throw light on the quantum measurement problem. \\

This paper is structured as follows: Section \ref{sec:foundations}
lays the foundations for the rest of the paper, setting out the
context for the discussion and presenting a basic viewpoint which is
then developed in the following sections.  A key aspect is the
proposal that higher level effective dynamics emerges out of lower
level dynamics. Section \ref{sec:linear} considers linear and
non-linear aspects of quantum theory, leading to some criteria for
when quantum physics will be valid, based on the essential linear
aspects of the theory.  Section \ref{sec:linear emerge} considers
when the requisite linearity can emerge at higher levels in the
hierarchy of complexity from lower level linear theories, and when
it cannot emerge. Section \ref{sec:contextual} looks at the converse
feature of how contextual effects from higher levels may influence
lower level dynamics, giving a number of examples of top-down
causation in the context of quantum physics. Section
\ref{sec:measure context} looks at the issue of state vector
reduction in the context of top-down causation from the local
physical environment. Section \ref{sec:classical emerge} looks at
implications of the discussion for the classical quantum cut and
Schr\"{o}dinger's cat. Section \ref{sec:conclusion} reviews the
viewpoint presented, and considers issues that arise from the
discussion as suitable subjects for
further investigation.\\

A major issue that arises out of the discussion in Section
\ref{sec:contextual} of top-down influences in physics is the origin
of the arrow of time. This is discussed in a companion paper
\cite{Ell11a}.

\section{Foundations}\label{sec:foundations} %
This section sets out the basic foundations for the rest of the
paper. Section \ref{sec:basic} sets out the basics of quantum
dynamics, Section \ref{sec:paradox} the elements of the measurement
problem, and Section \ref{sec:basis} sets out a basic standpoint
that underlies what follows.  Section \ref{sec:context} sets out the
context of the hierarchy of structure, and Section
\ref{level_relations} the viewpoint that both bottom-up and top-down
causation take place in this hierarchy.

\subsection{Basic dynamics}\label{sec:basic}
The basic expansion postulate of quantum mechanics
\cite{Mor90,Rae94,Ish95,GreZaj06} is that before a measurement is
made, the state vector $\ket\ps$ can be written as a linear
combination of unit orthogonal basis vectors
 \be
 \ket{\ps_1} = \sum_n c_n\ket{u_n(x)}, \label{wave}
 \ee
where $u_n$ is an eigenstate of some observable $\hat{A}$ (\cite{Ish95}:5-7).
The evolution of the system can be completely
described by a unitary operator $\widehat{U}(t_2,t_1)$, and so
evolves as

\be \ket{\ps_2} = \widehat{U}(t_2,t_1)\,\ket{\ps_1} \label{U1} \ee

\noindent Here $\widehat{U}(t_2,t_1)$ is the standard
evolution operator, determined by the evolution equation

\be i\hbar\frac{d}{dt} \ket{\ps_t} = \hat{H}  \ket{\ps_t}.
\label{evolution} \ee When the Hamiltonian $\hat{H}$ is time
independent, $\widehat{U}$ has the form (\cite{Ish95}:102-103)

\be \widehat{U}(t_2,t_1) = e^{-\frac{i}{\hbar} \hat{H}(t_2-t_1)}
\label{U2} \ee which is unitary (\cite{Ish95}:109-113):
\begin{equation}  \label{U3}
\widehat{U}\, \widehat{U}^\dag = 1.
\end{equation}

Applying this to (\ref{wave}) with $\widehat{U}(t_2,t_1) \ket{u_n(x)} =  \ket{u_n(x)}$ (an invariant basis) gives
\be
 \ket{\ps_2} = \sum_n C_n\ket{u_n(x)},\,\,\,\,\, C_n := \widehat{U}(t_2,t_1) c_n. \label{wave2}
 \ee

Immediately after a measurement is made at a time $t=t^*$, however,
the relevant part of the wavefunction is found to be in one of the eigenstates:
 \be
 \ket{\ps_2} = c_N \ket{u_N(x)} \label{collapse}
 \ee
for some specific index $N$. This is where the quantization of
entities and energy comes from (the discreteness principle): only
eigenstates can result from a measurement. The eigenvalue $c_N$ is
determined by the operator representing the relevant physical
variables, and hence is unrelated to the initial wave function
(\ref{wave}). The data for $t< t^*$ do not determine either $N$ or
$c_N$; they merely determine a probability for each possible
outcome (\ref{collapse}), labelled by $N$, through the fundamental
equation
 \be
p_N = c_N^2 = \langle e_N | \psi_1\rangle^2.  \label{prob}
 \ee
One can think of this as due to the probabilistic time-irreversible
reduction of the wave function
\be
 \begin{array}{c c c}
 \ket{\ps_1} = \sum_n c_n\ket{u_n(x)} \hspace{1.75 cm}
 \longrightarrow \hspace{1.5 cm}\ket{\ps_2} = c_N u_N(x)\nn\\
 Indeterminate  \hspace{2 cm} Transition  \hspace{1 cm} Determinate \\  \label{trans}
\end{array}
\ee
This is the event where the uncertainties of quantum theory become
manifest (up to this time the evolution is determinate and time
reversible). It will not be a unitary transformation (\ref{wave2})
unless the initial state was already an eigenstate of $\hat{A}$, in
which case we have the identity projection
 \be
 \ket{\ps_1} = c_N u_N(x) \hspace{1.75 cm}
 \longrightarrow \hspace{1.5 cm}\ket{\ps_2} = c_N u_N(x)
 \label{trans1}
 \ee
Hence it is unclear how this experimental result can emerge from the
underlying quantum theory, which leads to (\ref{wave2}) rather than
(\ref{collapse}). It is also unclear how classical behaviour can
emerge from the underlying quantum behaviour, which will generically
show (a) entanglement between different entities so that they do not
have distinct individual states, and (b) only probabilities of
different values of physical variables rather than specific
determinate values.\\

This discussion presents the simplest idealized case of a
measurement (\cite{Pen04}:542-549). More generally, one has
projection into a subspace of eigenvectors (\cite{Ish95}:136;
\cite{WisMil10}:10-12) or a transformation of density matrices
(\cite{Ish95}:137), or any other of a large set of possibilities
(\cite{WisMil10}:8-42), but the essential feature of
non-unitary evolution remains the core of the process.
Thus there is a  deterministic prescription for evolution of the quantum state
determining probabilities of outcomes of measurements, but
indeterminacy of the specific outcomes of those measurements, even if the
quantum state is fully known. Examples are radioactive decay (we
can't predict when a nucleus will decay or what the
velocities of the resultant particles will be), and the foundational
two-slit experiments (we can't predict precisely where a photon,
electron, neutron, or atom will end up on the screen
\cite{FeyHib65,GreZaj06}).\\

The fact that such unpredictable measurement events happen at the
quantum level does not prevent them from having macro-level effects.
Many systems can act to amplify them to macro levels, including
photomultipliers (whose output can be used in computers or
electronic control systems). Quantum fluctuations can change the
genetic inheritance of animals \cite{Per91} and so influence the course of
evolutionary history on Earth, and they have changed the course
of structure formation in the universe \cite{Ell06}. Thus quantum
implications are not confined to the micro realm.

\subsection{The measurement problem}\label{sec:paradox}
The measurement problem \cite{LonBau83,WheZur83,All11} or measurement paradox
(\cite{Pen04}:782-815) is a key issue for quantum theory. Measurement is the
location of the unpredictability of outcomes, consistent with
the quantum uncertainty relations (see e.g. \cite{Mor90,GreZaj06}); but it is not
an outcome of standard quantum dynamics, although it is crucial to the theory.\\

It is a fundamental aspect of quantum theory that the uncertainty of
measurement outcomes is unresolvable: it is not even in principle
possible to obtain enough data to determine a unique outcome of
quantum events \cite{FeyLeiSan65,Ish95,Pen89}. This unpredictability
is not a result of a lack of information: it is the very nature of
the underlying physics. This uncertainty is made manifest when a
measurement takes place, and only then - without measurements, there
is no uncertainty in quantum processes. Here we mean by a
\emph{measurement}, a process whereby quantum uncertainty is changed
into a definite classical outcome that can be recorded and examined
as evidence of what has happened; it is not necessary that an
observer actually takes any measurements. For example it happens
when a photon falls on a physical object such as a screen, a
photographic plate, or the leaf of a plant, and deposits energy in a
particular spot on the object. In more technical terms, it
generically occurs when some component of a general wavefunction
collapses to an eigenstate of an operator (eqn.(\ref{trans}) above).
And this is not a side effect in quantum theory: it is absolutely
central to its real world applications. As stated by Leggett,
\begin{quote}
``.. it is the act of measurement that is the bridge between the
microworld, which does not by itself possess definite properties,
and the macroworld, which does. .. the concept of measurement, prima
facie at least, is  absolutely central to the interpretation of the
quantum mechanical formalism'' (\cite{Leg91}: 87).
\end{quote}

However the process of determining experimental results --- a
measurement --- cannot be represented by the standard quantum state
evolution equations, such as the Schr\"{o}dinger and Dirac equations,
for those are predictable (they obey existence and uniqueness
theorems) and time reversible. They simply do not have the kind of
nature that can lead to an unpredictable result when the initial
state is fully known (\cite{Pen04}:530-533); but that is what
happens in quantum measurements, which do not obey linearity and
hence violate the superposition principle.\\

This is the \emph{measurement paradox}: the process of measurement
(\cite{Mor90}:80-102,491-556,591-619; \cite{Rae94}:53-62;
\cite{Ish95}:175-188; \cite{GreZaj06}:215-243; \cite{Pen89}:225-296;
\cite{WisMil10}:8-44) cannot be described by standard quantum
dynamics. Indeed, Leggett states it thus (\cite{Leg91}:87,89):

\begin{quote}
``the problem is that quantum mechanics absolutely forbids a
measurement to take place .....  in a nutshell, in quantum mechanics
events don't (or don't necessarily) happen, whereas in our everyday
world they certainly do''.
\end{quote}
Aharonov and Rohrlich, considering a Stern-Gerlach apparatus with
entangled spin states (\cite{AhaRoh05}:122), express it this way:
\begin{quote}
``Clearly, our treatment of quantum measurements is incomplete, we cannot leave the
measuring device in a superposition of states. But clearly quantum mechanics offers
no way to reduce a superposition of pointer states to a definite position.''
\end{quote}

Now this has been disputed by many, and alternative descriptions
have been proposed that try to get around this fundamental
limitation.\\

The \emph{\textbf{many-worlds view}}
\cite{Rae94,Ish95,Wal01,Wal08,Sauetal11} theoretically involves only
unitary processes; technically it is based on the idea of a relative
state, which involves a special basis of Hilbert space relative to
which the splitting occurs (\cite{Ish95}:157-159). This approach
arises out of assuming the Schr\"{o}dinger equation
(\ref{evolution}) applies consistently to the physical universe at
all scales, and taking the consequences seriously. However it will
be the contention of this paper that the application of that
equation is strictly limited (Section \ref{limits}). I will here
pursue the idea that some form of state vector reduction should be
taken seriously.\\

In the \emph{\textbf{consistent histories approach}} \cite{Har08,Wal08}, a Heisenberg
formalism is used with unitary evolution of the projection operators  but
non-unitary projections taking place to define a
branch state vector (\cite{Har08}: eqn.(A.2)). Thus even though the
operator evolution is unitary, this formalism does not get rid of
the need for non-unitary projection operators, and some prescription as to
when they should operate
(see also comments in \cite{All11}).\\

\emph{\textbf{Decoherence}} \cite{Ish95,Zur04,GreZaj06} does not
solve the problem either, as some claim. The measurement problem
involves two distinct steps. The first is the non-unitary
elimination of the off-diagonal terms of the density matrix
(decoherence). The result is a statistical ensemble. The second is
the projection of one particular eigenvalue from that ensemble.
Decoherence does not solve step 2. It effectively removes
entanglement (by diagonalising the density matrix), but the
diagonalised density matrix still does not determine or even
represent a unique outcome for a specific physical situation. But we
want a theory can that can at least \emph{describe} a specific
result for an individual entity, even if it can't causally predict
that unique outcome. The way theoretical physics underlies the real
world, including biology, must apply to unique
individuals as well as to statistical ensembles:
for ensembles are made of individual entities.\\

None of these proposed alternatives solves the measurement paradox
in a way that changes the fundamental lack of predictive capacities
of quantum theory. As far as real physical experiments are
concerned, what happens is described by the equations presented in
section \ref{sec:basic}; this is what has to be explained. It took
decades of theoretical exploration
and experimental work to verify that this is the way things work; those results are not in doubt,
and are what we have to deal with.\\

By contrast the \emph{\textbf{explicit collapse models}}, for
example those by Ghirardi, Rimini, and Weber \cite{BasGir03,Ghi07}
and Penrose \cite{Pen89,Pen04}, do offer a solution to the
problem. However I am here concerned with standard quantum theory;
I will propose (Section \ref{quest}) the possibility that such an
effective collapse mechanism might arise as a local top-down
effect from a measuring apparatus. I will not deal with
\emph{\textbf{hidden variable theories}}
\cite{WheZur83,Rae94,Ish95}, except to comment that
the same issues of interacting scales will occur in those theories too.\\

There are of course various
different approaches to ontology in quantum theory, with realist,
anti-realist and pragmatic viewpoints on offer \cite{Ish95}; I do
not intend to discuss those issues explicitly here, although my
tendency will be towards a realist viewpoint.

\subsection{A basic standpoint}\label{sec:basis}
To ground this analysis in reality, I will adopt the following
starting point for what follows:\\

 \textbf{BASIC PREMISE}: \emph{Individual Events Happen}. \\

 \noindent Each word is important:\\

\textbf{Individual}: Statistics is not enough. An ensemble of events
is made up of individual events. There is no ensemble if individual
events don't separately happen.

\textbf{Events}: Specific things occur. Universal laws describe
multifold possibilities of what might happen, but we experience
specific events in our own particular history.

\textbf{Happen}: They occur in time: they are about to occur, they
occur, then they have occurred. Uncertainty about what might occur
changes to the certainty of what has occurred.\\

\textbf{What is the evidence for this statement?} Apart from the
overwhelming evidence from everyday life, every single physics
experiment is proof it is true! - we plan experiments, carry them
out, analyse the results, publish them. Each experiment is an individual
event that occurs at a particular time and place in the history of the universe.
Science would not be possible if this were not the case.
\\

\textbf{This is true at every level of the hierarchy of structure
and complexity}, summarized in Table 1 below.

\noindent At the \textbf{macrolevel}:

 - the universe evolves, structures form,

 - stars explode, planets move round the
 sun,

 - objects fall to the surface of the earth, birds fly.

\noindent At the \textbf{micro level}:

 - electrons in atoms change energy levels and emit photons, or absorb photons and change energy levels

- particles are emitted, go through slits, get scattered, impact on
screens

 - photons are emitted, go through polarizers, get scattered,
are detected

- entanglement and decoherence take place.\\

This takes place irrespective of whether we know about it or not.
Observers are not necessary for things to happen! Events (e.g.
nucleosynthesis) took place in the early universe long before any
physical observers existed. In the absence of observation the wave
function evolves unitarily, so in this case there is no clear
meaning to before and after.  Hence effective ``observation'' (i.e.
 collapse of the wave function) takes place all the time, because
definite outcomes are occurring all the time, whether actually
observed or not \cite{Ell11a}.\footnote{What does it mean to say an
event ``has happened'' if the system of interest is, say, a particle
passing through a two-slit system in the absence of observation?
Quantum mechanics forbids us from saying that (for example) the
particle ``really did'' go through slit A, wave function collapse
takes place --- which can be much later. This can result in a
delayed ``passage of time'' in an evolving block universe, see
\cite{EllRot10} for a discussion.} Hence we do not need to involve
consciousness in quantum theory foundations (ontology), although it
is relevant to epistemology. Experimenters can make things happen,
but they carry on happening whether observers exist or not.\\

The \textbf{implication} is that any complete theory of causation in
the physical world, whether deterministic or not, must in some sense
explain what is happening in specific instances both for inanimate
matter and for life, otherwise it will be an incomplete explanation
of the real world. It may not be able to explain \emph{why} they are
happening (it may assume there is some irreducible randomness that
acts as as an effective cause, for example), but it must at least be
able to describe \emph{that} they are happening, as is shown by
experiment, i.e. it must represent the fact that reduction to
eigenstates takes place. Quantum theory predictions of energy
levels, scattering angles, and statistics of interactions in general
are of course sound testable physics, and it is a major success that
they are correctly predicted. But they are not a complete theory of
causation and events: they don't even account for when specific
events occur. They must be supplemented by some standpoint on
\emph{when} state vector reduction occurs in order to relate
adequately to the macro world, even if they don't uniquely predict
the outcome. Simply to pragmatically ignore the problem is no
resolution \cite{All11}. The implications for the nature of time are
considered in \cite{Ell11a}.

\subsection{The context: the hierarchy of the structure and causation}\label{sec:context}
The context in which this all occurs is the hierarchy of structure
and causation. In simplified form, this is as set out in Table 1.
This Table gives a simplified representation of this hierarchy of
levels of reality as characterized by corresponding academic
subjects, with the natural sciences on the left and the life
sciences \cite{CamRee05} on the right. On both sides, each lower
level underlies what happens at each higher level in terms of
structure and causation. Note that there is no correlation between
the left and the right hand columns above the level of chemistry, as
emergence and causation is quite different in the two cases; but the
first four levels are identical (life emerges out of physics!). On
the left hand side higher level correspond to larger scales (each
level is the encompassing domain or environment of the next lower
level); on the right hand side goals and intentions are relevant, so
that is what the higher levels refer to. \footnote{The labels
``higher'' and ``lower'' are sometimes contested, but seem to
provide a useful framework if they are defined in this way.}\\

\begin{center}
\begin{tabular}{|l|l|l|}\hline
 Level 10: &  Cosmology &  Sociology/Economics/Politics\\ \hline
  Level 9: &  Astronomy  &  Psychology \\ \hline
  Level 8: &  Space science &  Physiology\\ \hline
  Level 7: &  Geology, Earth science &  Cell biology \\ \hline
  Level 6: &  Materials science &  Biochemistry \\ \hline
  Level 5: &  Physical Chemistry &  Chemistry \\ \hline
  Level 4: &  Atomic Physics &  Atomic Physics \\ \hline
  Level 3: &  Nuclear Physics &  Nuclear Physics \\ \hline
  Level 2: &  Particle physics &  Particle physics\\ \hline
  Level 1: &  Fundamental Theory &  Fundamental Theory  \\
  \hline
\end{tabular}\\
\end{center}

{}\\ \textbf{Table 1:} \emph{The hierarchy of structure and
causation for
inanimate matter (left) and for life (right). 
 For a more detailed description 
 see }http://www.mth.uct.ac.za/$\sim$
 ellis/cos0.html.
\\

\noindent Implicit in this discussion is the view that the elements
at each of the levels characterized by this table, except perhaps at
the quantum level \cite{GreHorZei89,Lapetal11}, can be regarded as
existing \cite{Edd28,Ell04}. A table exists, even though it is made
of atoms, which also exist, even though they are made of electrons,
protons, and neutrons; and of course the same applies to animals and
people. This view too is needed in order that science makes sense.
If an experimenter does not exist, then
experiments are not possible.\\

Quantum Mechanics is applicable at the lower levels, but apparently
not at the macrolevels except under very restricted circumstances
 - for example superconductivity, Bose-Einstein condensations, lasers, and the
extraordinary recent quantum entanglement experiments over many
kilometers. It is not apparent under ordinary every day
circumstances at the macro level (which is why quantum dynamical
principles are not obvious to us). Hence experimenters talk about
the classical/quantum cut, or Heisenberg cut (\cite{WisMil10}:15),
necessary for them to analyze their experiments.

\subsection{Inter--level relations}\label{level_relations}
The higher and lower levels are related to each other
because the higher levels are based in the lower levels. To characterize causation in
this hierarchical context, it is useful to consider causation as proceeding in both a
bottom-up and a top-down manner \cite{Ell08}.

\subsubsection{Bottom-up Effects} A major theme of physics is that
causation occurs from the lower to the higher levels of the
hierarchy, leading to the emergence of structure and complexity. A
feature that occurs here is the \emph{coarse-graining} of lower
level variables (e.g. particle states) to give higher level
variables (e.g. density and pressure) \cite{AloFin71}, accompanied
by a conversion of useful energy to non-usable energy when some
energy is hidden in lower level variables, and hence not available
to higher levels. This is the source of entropy growth and of
effective non-conservation of energy at higher levels through
friction and
other dissipative effects.\\

But there are limits as to how far this bottom-up process of
explanation can be carried out: physics per se cannot explain
economics or psychology, for example. \emph{More is different}, as
famously stated by Anderson \cite{And72}; emergence of complexity
takes place where quite different laws of behaviour hold at the
higher levels than at the lower levels \cite{Kau93,Gel94}. Laughlin
has elaborated how some higher level effects can only be understood
in terms of variables expressed in terms of higher level concepts
\cite{Lau00}. In particular, the linearity of lower level laws gets
replaced by the complexity of non-linear interactions at higher
levels, without which life could not come into existence.

\subsubsection{Top-down effects} In addition to bottom-up causation,
\emph{contextual effects} occur whereby the upper levels exercise
crucial influences on lower level events by setting the context and
boundary conditions for the lower level actions. This is related to
the emergence of effective laws of behaviour at higher levels that
enable one to talk of existence of higher level entities in their
own right. They then play an effective role not only at their own
levels, but also influence the lower levels by
setting the context for their action \cite{Ell08}.\\

This idea of top down action in physics goes back at least to Ernst
Mach in his work on Mach's principle and the origin of inertia
(\cite{EllSci72,Ell02};\cite{Sil01}:58-61), which strongly
influenced Albert Einstein in developing general relativity theory
and his static universe model. It is crucial in ideas about the
origin of the arrow of time
\cite{WheFey45,EllSci72,Dav74,Ell02,Pen04,Car10}; nice popular
discussions of how top-down effects may take place from the universe
to local physics are given in \cite{Sci59,Cho10}. I will make the
case that \emph{top-down influences play a key role in relation to
how quantum theory works}, particularly as regards both decoherence
and state preparation. It is possible this line of thought can
illuminate the way quantum measurement takes place.

\section{Linearity, non-linearity, and quantum theory}\label{sec:linear}
The key issue I now focus on is how linearity and non-linearity in
QM relate to each other. Section \ref{sec:linear1} considers the
essential linearity of Quantum Mechanics (QM), and Section
\ref{sec:nonlinear} its essential non-linearities. Section
\ref{sec:non-linear1} looks at allowed non-linearities, and Section
\ref{main} at when we may expect QM to be valid, in the light of the
above sections.

\subsection{The essential linearity of QM}\label{sec:linear1}
Linearity is at the core of quantum theory. Ghirardi
\cite{Ghi07}  states it thus:
\begin{quote}
``Let us recall the axiomatic structure of quantum theory: 1. States
of physical systems are associated with normalized vectors in a
Hilbert space, a complex, infinite-dimensional, complete and
separable linear vector space equipped with a scalar product.
Linearity implies that the superposition principle holds: if
$\ket{f}$ is a state and $\ket{g}$ is a state, then (for $a$ and $b$
arbitrary complex numbers) also $\ket{K} = a\ket{f} + b\ket{g}$ is a
state. Moreover, the state evolution is linear, i.e., it preserves
superpositions: if $\ket{f(t)}$ and $\ket{g(t)}$ are the states
obtained by evolving the states $\ket{f(0)}$ and $\ket{g(0)}$,
respectively, from the initial time $t=0$ to the time $t$, then
$a\ket{f(t)} + b\ket{g(t)}$ is the state obtained by the evolution
of $a\ket{f(0)} + b\ket{g(0)}$. Finally, the completeness assumption
is made, i.e., that the knowledge of its state vector represents, in
principle, the most accurate information one can have about the
state of an individual physical system.''
\end{quote}
This linearity is central to
\begin{itemize}
  \item the superposition principle for
quantum states: (Dirac \cite{Dir58}:12-18, Isham \cite{Ish95}:4,11),
see also \cite{Mor90,Rae94,Ish95,GreZaj06}, leading to
  \item interference between quantum entities as in the 2-slit experiment(\cite{FeyHib65}:4-6),
  \item development of entanglement (\cite{Ish95}:148-149),
  \item linearity of the wave function (\cite{Ish95}:15-16), hence
  \item expansion of the wave function in terms of a basis (\cite{Dir58}:53-67, \cite{FeyHib65}:86-87),
  \item thus it is the reason that wave functions live in a vector space
  \cite{Ish95}:19-20) and so is why a Hilbert space formalism is
  suitable for quantum theory (\cite{Dir58}:40, \cite{Ish95}:19-35, 71, \cite{Pen04}:530-538).
 \item It is based in the way the amplitude is linear sum over paths (\cite{FeyHib65}:6,19,29), and hence
  \item is embodied in the Schrodinger and Dirac equations, both of the form
  (\ref{evolution}), leading to
 \item unitary transformations (\cite{Dir58}:103-107; \cite{Ish95}
  113-115).
  \item It occurs when scattering of identical particles takes place (\cite{FeyLeiSan65}:(4.1),(4.2)), and so
  \item underlies the unitarity of the S-matrix (\cite{ItzZub80}, 166-167; \cite{Sch68}, 307-319),
  \item as well as Bose-Einstein and Fermi-Dirac statistics (\cite{FeyLeiSan65}:4-3 to 4-15), because
  permutation of states (\cite{Dir58}:207-216) is a linear operation.
  \end{itemize}
Consequently, it is crucial in applying quantum theory to physics and chemistry \cite{Rio05}.\\

In more detail: quantum theory is applicable when the evolution of the state vector is linear. It takes the
form (\ref{U1}):
\begin{equation}  \label{eq:U4}
 \ket{\Psi} \rightarrow \ket{\Psi'} = \widehat{U} \, \ket{\Psi}
\end{equation}
where the operator $\widehat{U}$ is linear:
\begin{equation}  \label{eq:L}
 \forall \,a, \,b,\ket{\Psi},\ket{\Phi}:\,\, \widehat{U}\left(a \ket{\Psi} + b \ket{\Phi}\right) =
a\widehat{U}\ket{\Psi} + b \widehat{U} \ket{\Phi}.
\end{equation}
Because the norm of $\ket{\Psi}$ is preserved, $\widehat{U}$ is a unitary matrix (\ref{U3}).
It is given by (\ref{U2}) in terms of the Hamiltonian; consequently
energy is preserved
\begin{equation}  \label{eq:U3}
H\ket{\Psi} = E \ket{\Psi} \Rightarrow\,\, E \,= \,const
\end{equation}
because anything that commutes with $H$ is a constant, and $H$
commutes with itself.\\

This relates to the Feynman path integral for particle motion in the
following way (\cite{FeyHib65}:26-29): the probability $P(b,a)$ for
a particle to go from a point $x_a$ at time $t_a$ to the point $x_b$
at time $t_b$ is the absolute square of an amplitude $K(b,a)$ to go
from $a$ to $b$ :
\begin{equation}
P(b,a) = |K(b,a)|^2. \label{eq:p1}
\end{equation}
The amplitude is a sum of contributions from each path between
events $a$ and $b$:
\begin{equation}
K(b,a)  = \sum_{paths} \phi[x(t)] \label{eq:p2}
\end{equation}
 where the contribution of each path has a phase proportional to the
 action $S$:
\begin{equation}
\phi[x(t)] = A e^{(i/h) S[x(t)]}\label{eq:p3}
\end{equation}
and the action is the path integral
\begin{equation}
S[x(t)] = \int_{t_a}^{t_b} L(\dot{x},x,t) dt \label{eq:p4}
\end{equation}
where $L(\dot{x},x,t)$ is the Lagrangian of the system
(\cite{FeyHib65}:26). Linearity follows from this definition of $S$
and form of $L$, because by the way integrals are defined, for any
time $t_c$ between $t_a$ and $t_b$, the action along any path
between $a$ and $b$ can be written
\begin{equation}  \label{eq:L1}
S(b,a) =  S(b,c)\,+ \, S(c,a)
\end{equation}
where $c$ is a point for which $t=t_c$ (\cite{FeyHib65}:36,76). This
underlies the key property of path integrals:
\begin{equation}  \label{eq:P1}
K(b,a) =  \int_{x_a}^{x_b} K(b,c) K(c,a) dx_c
\end{equation}
which follows on integrating over all values $x_c$
(\cite{FeyHib65}:37), leading to the wave function integral equation
\begin{equation}  \label{eq:w1}
\psi(x_b,t_b) =  \int K(x_b,t_b;x_c,t_c) \psi(x_c,t_c) dx_c
\end{equation}
(\cite{FeyHib65}:57), which is linear in $\psi$, even if the
kernel $K$ is non-linear.

\subsection{Essential non-linearities of QM}\label{sec:nonlinear}
However there are also two essential non-linearities in quantum mechanics.\\

The first is the way probabilities are derived as squares of the wave function
(eqn.(\ref{prob}); \cite{Ish95}: eqn.(5.29)) or equivalently  as squares of amplitudes
(eqn.(\ref{eq:p1}); \cite{FeyHib65}:29).
It is this non-linearity that lies
at the heart of the difference between classical and quantum statistics (\cite{FeyLeiSan65}:1-1 to 1-10,
\cite{Ish95}:11-14).
This carries over to the way expectation values are derived for any operator (\cite{Dir58}:45-48;
\cite{Ish95}:83-84):
\begin{equation}  \label{eq:expect}
\langle A\rangle_\psi \,= \,\bra{\psi} A \ket{\psi},
\end{equation}
which is quadratic in the wave function (and hence non-linear).
This non-linearity is compatible with the unitary evolution (\ref{evolution}), indeed it is essential to its
interpretation, and underlies the way a normed vector space representation of quantum
 probabilities makes sense (\cite{Ish95}:13-14).\\

The second essential non-linearity is state vector projection (\ref{trans}), which is not compatible
with the unitary evolution (\ref{evolution}), (\ref{eq:U4}), see Section \ref{sec:paradox}. It is essential because
probabilities of outcomes depend on individual events happening; statistics of measurements can only
emerge from specific individual measurements that have separately occurred. We return to the measurement
issue in Section \ref{sec:measure context}.

\subsection{Allowed and non-allowed non-Linearity in QM}\label{sec:non-linear1}
But additionally, non-linearities allow quantum theory to describe many non-linear effects, for example
those expressed in Feynman diagrams. How can that
happen in a way compatible with what has been said here about linearity (Section \ref{sec:linear})?
Basically, both through linear systems being imbedded in non-linear
environments in such a way that linearity is locally preserved for
subsystems of the whole --- the linearity of the subsystem is not interfered with by the environment  --- and through
approximation methods involving higher and higher orders in a perturbation series.\\

Specifically, we can have arbitrarily complex behaviors in the Lagrangian, but the
probability amplitudes and wave function must be affected in a
linear way by the time evolution. Thus non-linearity in systems obeying
the QM relations can arise in two ways:\\

\textbf{Firstly} via the Lagrangian $L(\dot{x},x,t)$ in the action
(\ref{eq:p4}). For example for a particle of mass $m$ subject to a
potential energy $V(x,t)$

\begin{equation}  L = T - V(x,t), \,\,\,T :=\frac{1}{2}\,m\,\dot{x}^2 \label{L}
\end{equation}
Apart from the non-linearity in the kinetic energy, the potential
can be arbitrarily non-linear and non-linear behavior can result,
but can sometimes be soluble. The Thirring model \cite{Tir58} for
example is an exactly solvable quantum field theory which describes
the self-interactions of a Dirac field in two dimensions. The matter
Lagrangian is therefore of necessity non-linear; but the equation
of motion for the wave function is a linear p.d.e.\\

 \textbf{Secondly}
through the expansion of the exponential in (\ref{eq:p2}):

\begin{equation}
\exp(iS ) = \sum_n \frac{(iS)^n}{n!} = 1 + iS - \frac{1}{2!}S^2
- \frac{i}{3!}S^3 +\dots
\end{equation}
The complexities of Feynman diagrams arise from this series of
nonlinear terms $S^n$ (\cite{FeyHib65}:120-125), where additionally
(by (\ref{eq:p4}), (\ref{L})), they themselves are non-linear terms: $S =
S(V,T)$. Non-linearities result from the many different interactions
represented by the terms in this expansion. But still in these cases
the action of $U$ must be linear in
 $\ket{\Psi}$  as in (\ref{eq:U4}), (\ref{eq:L}) (Dirac), hence also as
 in (\ref{eq:w1}) (Feynman), in order to be compatible with the foundations
 of quantum physics.\\

However nonlinearities can also arise in ways that are incompatible
with these linear foundations: there is no guarantee that higher
level emergent behaviour will obey (\ref{eq:U4}), (\ref{eq:L}),
indeed generically it won't do so. Equivalently, there are
situations where a path integral formalism
(\ref{eq:p1})-(\ref{eq:p4}) is simply not applicable (at
the chosen level of description). I will give examples below.\\

\textbf{The implication} is that in order that the unitary quantum
mechanics formalism can be applicable, one must select a subsystem
of the complex interacting whole where the unitary aspect
(\ref{eq:U4}) is true. This is what occurs when one focuses on the
relations between elementary particles, for example, and when one
constructs superconducting systems or lasers or Bose-Einstein
condensates. But these are very special cases, as is shown by the
care one has to take in constructing such systems. It is not
possible to do this for generic complex
systems - or even for some quite simple macro systems. \\

Before considering
this in the next section, we need to consider three queries that might undermine
this claim.\\

Firstly, why is the above argument not vitiated by the existence of
\textbf{nonlinear versions of the Schr\"{o}dinger equation} (`NLS')? \cite{FisKriSof11}.
The problem is that since the equations themselves are nonlinear, the solutions can't be
superposed in general. There are exceptions: plane wave solutions
exist for the nonlinear Schr\"{o}dinger equation \cite{Pol04}, and these can be
superposed in special cases, but this is not possible generically,
e.g. you can't superpose two plane waves with different propagation
directions. NLS equations don't describe the evolution of a general
quantum state, because they only obey the superposition principle
for very special cases; hence they do not describe generic
situations of either interference or entanglement, which are central
to quantum theory, rather they are classical field equations for
fiber optics and water waves. When canonically quantized,
the NLS equation describes bosonic point particles with
delta-function interactions, and the related Gross-Pitaevskii
equation describes the ground state of a quantum system of identical
bosons. Thus they deal with very particular physical cases, not
related to the context of generic emergence of higher level systems I consider here.
In the latter case the non-linear Gross-Pitaevskii equation is
not an equation for a quantum mechanical wave function, even though
it is often called the wave-function of the condensate; it is an
equation for a classical field having the meaning of an order
parameter.\\

Secondly, what about \textbf{linear solutions to other non-linear
equations that might describe physical behaviour}? There are linear
families of solutions to some special non-linear equations such as
the Sine-Gordon equation; could they not be used in a theory that
satisfies the linearity requirements discussed above?
Similar comments apply to those above: this is not
possible for a theory that covers generic physical conditions, because
in the case of these equations, this linearity only holds for
special initial conditions; but a general physical theory must apply
to very general initial conditions. The merit of linear solutions is
that (as is shown by Fourier analysis) they can represent almost
any initial conditions: so solutions to the linear equation (\ref{evolution})
are not restricted to specific kinds of initial states.\\

Thirdly what about thinking of \textbf{quantum theory as a theory of
perturbations?} Almost any system can be described in a perturbation
series, where the linear terms will dominate the dynamics in many
cases: it's behaviour will be linear for all practical purposes
(FAPP), even though the system as a whole may be highly non-linear.
Many quantum phenomena can indeed be regarded in this way. So
perhaps we can regard quantum theory as a theory of perturbations
which can be applied
locally to almost any situation, even if it cannot be applied globally.\\

This view  has merit.
However in some cases, there is no linear perturbation theory, in the
sense demanded by QM, as a good approximation to higher level dynamics.
I will give examples later on. In any case this formulation makes it clear this will
only cover restricted physics situations: it will not apply when the non-linearities really matter.

\subsection{When is quantum mechanics valid?}\label{main}
The conclusion is that QM centrally implies linearity; so attempts to
extend quantum physics to macro objects requires selecting a linear
subsystem from nonlinearity (hence it has in particular to be
shielded from environmental noise). But there may be cases where
this is not possible. Leggett states (\cite{Leg91}: 98),
\begin{quote}
``It is quite conceivable that at the level of complex, macroscopic
objects the quantum mechanical superposition principle simply fails
to give a correct account of the dynamics of the system''.\end{quote}
If this is the case, then higher-level emergent dynamics are the
true determinants of what happens at macroscopic levels, and quantum
physics per se is not applicable as an effective theory determining outcomes at those
levels.\\

Why should one think this to be the case? Superposition is a
consequence firstly of the fact that the quantum state lives in a
vector space, with its linear structure appropriate to probability
measures, and secondly of the fact that the evolution equations for
the quantum state vector are linear first-order differential
equations in time, and so respect this linear structure. However,
\emph{inter alia} we want to consider how causality works in the
case of biological systems (the right hand column of Table 1). Such
complex systems are based in networks of interactions (such as gene
networks, protein networks, neural networks, brain circuits) that
involve non-linear structural and causal relations between
constituent elements \cite{Pea89,KanSchJes00,Alo07}, so
superposition surely would not be expected to hold in them. Note
that as discussed above, ordinary quantum theory allows a certain
degree of non-linearity in that it allows non-linear potentials to
occur in the linear time-development equations. It is the linearity
of the time development equations that matters there, and that is
what is violated in generic networks: the higher-level structure of
the system introduces non-linearities such as network motifs into
the dynamics \cite{Alo07}.

\subsubsection{The central proposal}\label{main1}
Accordingly, we can make the central proposal of this paper as follows:\\

\textbf{Proposal: Nature of physical reality}
\begin{enumerate}
  \item \emph{\textbf{Combinatorial structure}: Physical reality is made of linearly behaving components
  combined in non-linear ways}.
  \item \emph{\textbf{Emergence}: Higher level behaviour emerges from this lower level structure}.
  \item \emph{\textbf{Contextuality}: The way the lower level elements behaves depends on the
  context in which they are imbedded}.\footnote{As mentioned later,
  this use of the term ``contextuality'' here is not the same as the
  rather specific way it is sometimes used in discussions
  on the foundations of quantum theory (see \cite{Zei99,Kiretal09} and
  references therein). The use in this paper is carefully explained below
  (Section \ref{topdown}).}
  \item \emph{\textbf{Quantum Foundations}: Quantum theory is the universal foundation of what happens,
through applying locally to the lower level (very small scale)
entities at all times and places.} \item  \textbf{Quantum
limitations}: \emph{The essential linearity of quantum theory
cannot be assumed to necessarily hold at higher (larger scale)
levels: it will be true only if it can be shown to emerge from the
specific combination of lower level elements}.
\end{enumerate}
\noindent The last statement is an implication of the previous ones.
It is something like a macrolevel superselection rule, which is not implied
by the decoherence mechanism.\\

Thus there are limits on quantum theory, which won't apply
at higher levels when the context creates non-linearity at those
levels in such a way that superposition is impossible; hence this is
a route to creation of macro objects not subject to QM. This further
suggests that quantum theory applications dealing with essentially
non-linear phenomena do so
  by 
  introducing classical elements into the experimental
  description (i.e. invoking a ``quantum-classical cut''  (\cite{WisMil10}:15)). This is in
  accord with the Copenhagen interpretation of quantum
theory (\cite{WheZur83}; \cite{Ish95}:132) and the way classical
apparatus is routinely invoked in quantum experimental setups (see
e.g. \cite{BrePet06}:93,327; \cite{Leo10}:108,110,122;
\cite{WisMil10}:77,84,93).\\

In the following, the concept of the combinatorial structure of
matter will be present throughout.  The theme of emergence is picked
up in Sections \ref{emergeh} and \ref{emergeq} in general, in
Section \ref{sec:linear emerge} as regards linear systems, and in
Section \ref{sec:classical emerge} as regards classical systems. The
theme of contextuality
  is followed up in Section
\ref{sec:contextual} in general, and in relation to the measurement
issue specifically in Section \ref{sec:measure context}. The way
quantum foundations underlie classical systems is discussed in
Sections \ref{cut} and \ref{sec:conclusion}.

\subsubsection{The emergence of higher level behavior}\label{emergeh}
As a preliminary, we consider how higher level behavior relates to lower
level behavior in two adjacent levels in the hierarchy of complexity (Diagram 1).
As stated above, the fundamental viewpoint will be that the
higher level behavior emerges from that at the lower levels.
\begin{center}
\begin{tabular}{|l|c|l|c|}\hline
   Level $N+1$: & Initial state $I$ &  \emph{Higher level theory} $T$: $\Rightarrow$ & Final state $F$ \\ \hline
    & $\Uparrow $ &  \emph{\textbf{Coarse grain}} & $\Uparrow $ \\ \hline
  Level $N$:  & Initial state $i$ &  \emph{Lower level theory} $t$: $\Rightarrow$ &  Final state $f$ \\ \hline
\end{tabular}\\
\end{center}

{}\\ \textbf{Diagram 1:} \emph{The emergence of higher level
behaviour from lower level theory. Coarse-graining the action of the
lower-level theory results in an effective higher level theory}.\\

The dynamics of the lower level theory maps an initial state $i$ to
a final state $f$. Coarse graining the lower level variables, state
$i$ corresponds to the higher level state $I$ and state $f$ to the
higher level state $F$; hence the lower level action $t: i
\rightarrow f$ induces a higher level action $T: I \rightarrow F$. A
\emph{coherent higher level dynamics} $T$ emerges from the lower
level action $t$ if the same higher level action $T$ results for all
lower level states $i$ that correspond to the same higher level
state $I$ \cite{Ell08}, so defining an \emph{equivalence class} of
lower level states that give the same higher level action
\cite{AulEllJae08} (if this is not the case, the lower level
dynamics does not induce a coherent higher level dynamics, as for
example in the case of a chaotic system). Then on coarse graining
(i.e. integrating out fine scale degrees of freedom), the lower
level action results in an emergent higher level dynamics: the
effective theory at the higher level. Three key points follow.

\begin{quote}
\emph{\textbf{EM1: Non-Commutation}: coarse-graining and dynamical
action do not commute in general},
\end{quote}
inter alia because a great deal
of information is hidden in the higher level view, and also because
\begin{quote}
\emph{\textbf{EM2: Essential higher level variables}: not all effective higher level variables can be derived by coarse
graining} \cite{Ell08}.
\end{quote}
Consequently
\begin{quote}
\emph{\textbf{EM3: Emergent dynamics}: the effective higher level dynamics will in general
not be the same as the lower level dynamics} \cite{And72}.
 \end{quote}

\noindent Here are
some examples:
\begin{itemize}
  \item \textbf{E1: Statistical physics} The underlying atomic theory
  leads to the macroscopic gas laws, thermodynamics, and thermal properties of
  gases (\cite{AloFin71}:434-518). There is no similarity
  between the underlying theory and the emergent theory, except that concepts of mass, energy, and momentum
conservation apply at both levels.
  \item \textbf{E2: Electrodynamics} The process of coarse graining leads to the polarization density of
a polarized medium (\cite{Uma89}:343-349), where the electric
field $\textbf{E}$ is a coarse-grained version of the microscopic
electric field $\textbf{e}$, and the displacement vector
$\textbf{D} = \textbf{E}+ 4\pi\textbf{P}$ includes a polarization
term $\textbf{P}$ representing coarse-grained dipole terms
(\cite{Jac67}:103-108). The fields  $\textbf{D}$ and $\textbf{E}$
are related by a polarization tensor $\epsilon_{ij}$ such that
$D_i = \epsilon_{ij}E_j$. The tensor $\epsilon_{ij}$ depends on
the micro structure of the medium; in an isotropic medium,
$\epsilon_{ij} = \epsilon \delta_{ij}$ (using Cartesian tensors);
in an anisotropic medium this is not the case. The coarse grained
version of Maxwell's equations gives the divergence of
$\textbf{D}$ and curl of $\textbf{E}$, so a modified version of
the microscopic equations emerges. The emergent theory is largely
similar to the underlying theory.
  \item \textbf{E3: Gravitational theory} Coarse graining leads to
backreaction effects modifying the coarse grained Einstein equations
\cite{Ell84}, which can in principle significantly affect the macro
dynamics. However in the context of current cosmology, these are
very small effects \cite{Claetal11}: the emergent theory is very
similar to the underlying theory.
  \item \textbf{E4: Physics to Chemistry} The interactions of
  Fermions leads through the Fermi exclusion principle to the nature
  of the hydrogen atom (\cite{AloFin71}:109-148) and the electronic structure of atoms (\cite{AloFin71}:158-176)
  and so the periodic table \cite{Pau60,Atk94}; the nature of the chemical bond emerges from physics \cite{Pau60,Atk94}.
  There is no similarity between the underlying
  theory and the emergent laws.
  \item \textbf{E5: Chemistry to Microbiology and Life} The complex
  modular hierarchical structure of life emerges from the underlying
  physical and chemical laws \cite{CamRee05}. There is no similarity between the underlying
  theory and the emergent behaviour,except that concepts of mass and energy  balance apply at both levels.
\end{itemize}
In most cases, the underlying theory leads to a higher level theory
characterizing quite different behaviour (after all, that is the
essential content of Table 1).

\subsubsection{The emergence of higher level quantum behavior}\label{emergeq}
For quantum like behaviour to emerge at a higher level, one needs to select
subsystems of the emergent whole where interference and entanglement are possible.
When is this possible?
Firstly,
\begin{quote}
\textbf{LSS: Linear state space}: the relevant variables must live
in a linear space,
\end{quote}
A vector space structure for a state space can be natural even
mandatory, even with non-linear equations of evolution. Think of
non-linear wave equations eg water!: the waves form a vector space
under pointwise addition of functions.\footnote{I thank Jeremy
Butterfield: for these comments.} But there are plenty of (one
real-parameter groups of) non-linear maps on a vector space eg a
Hilbert space. So one can have superposability of states, but the
dynamics can fail to preserve a given superposition. One must avoid
this (cf. the quote from Ghirardi in Section \ref{sec:linear1}), so
the second requirement is
\begin{quote}
\textbf{LE: Linear evolution}: the relevant dynamical evolution must
be linear (a special case being unitarity), hence respects the
linear state space structure).
\end{quote}
Then the probability amplitude evolves linearly,  so we need to
find linearly behaving subsets of complex systems (possibly emerging
as collective modes of lower level components). Inter alia this means we must
\begin{itemize}
  \item \textbf{ L1:} restrict them in phase space terms, so that they remain in a linearly behaving domain of phase space
   (which will always be limited in both position and momentum terms, as eventually non-linearities will occur
    for larger size and energy scales).
  \item \textbf{ L2:} shield them from noise and interfering effects (so they must be isolated from the environment),
  \item \textbf{ L3:} restrict internal noise generation (so they must be cold),
  \end{itemize}
It is very difficult to attain such a situation on a macroscale, or
even a nano scale. Milburn (\cite{Mil97}:94-95) states this as
follows, in regard to electrons in a crystal structure:
\begin{quote}
    ``\emph{While we can carefully model the effect of the regular array of
    ions in the lattice, we have no knowledge at all of the details
    of the defects and impurities. Furthermore, at a finite
    temperature the ionic cores are wobbling around in a random way
    which we can only describe statistically. Were it not for these
    complications, we could use Schr\"{o}dinger's equation to
    assign probability amplitudes for an electron ....''}

    \emph{``There are two things we can do to prevent phase-destroying collisions. We can try and make
    ultra pure samples in which the defects and impurities are carefully
    controlled. This is exactly what is done, and indeed the artificial crystals grown to
    form such devices are probably the most pure and perfect artificial constructions ever made.
    The only way to reduce the effect of random lattice vibrations
    is to cool the devices.Typically liquid helium temperatures are
    used, a few degrees above absolute zero .... Quantum nanodevices are very cold, extremely tiny,
    near-perfect electrical devices}.''
\end{quote}
  This illustrates why we do not expect quantum behaviour to often emerge in a solid
  in a macro-context. It is easier to satisfy these conditions with
  photons, as in the case of quantum optics devices \cite{Leo10}.
  To investigate this further, it is useful to consider the variety of
examples where linearity can emerge at higher levels. I will
  first give some examples where this is possible (Section
\ref{sec:linear poss}), and then some where it is not (Section
\ref{sec:not linear}, which picks up on Point 5 in Section
\ref{main1}).

\section{Emergence of linearity}\label{sec:linear emerge}
Hence the issue is, under what conditions can an emergent higher
level behaviour resulting from low level quantum theory still be
described by the quantum theory laws of behaviour? (Diagram 2).\\

\begin{center}
\begin{tabular}{|l|c|l|c|}\hline
   Level $N+1$: & Higher level theory & & Emergent Theory ?\\ \hline
    & $\Uparrow $ &  \emph{\textbf{Coarse Grain}} & $\Uparrow $\\
  \hline Level $N$:  & Underlying theory & &  Quantum Theory  \\ \hline
\end{tabular}\\
\end{center}

 {}\\ \textbf{Diagram 2:} \emph{The emergence of higher level
behaviour from the underlying quantum theory}.\\

Suppose higher levels have effective laws valid at that level that
are emergent from the actions of lower level laws. Then behavior at
a higher level $N+1$ emerges from that at the lower level $N$.
Suppose the laws of quantum physics hold at level $N$ with
Hamiltonian $H_N$. Then three possibilities arise:
\begin{enumerate}
  \item \textbf{Case 1}: Quantum theory remains valid at level $N+1$ with the same Hamiltonian
  as at level $N$, i.e. $H_{N+1} = H_N$. Energy is conserved at level $N+1$
  \item \textbf{Case 2}: Quantum theory remains valid at level $N+1$ with the a different  Hamiltonian
  than at level $N$, i.e. $H_{N+1} \neq H_N$. Energy is conserved at level $N+1$.
  \item \textbf{Case 3}: Quantum theory is not valid at level $N+1$:  there is no Hamiltonian
 description applicable at that level: the evolution is not unitary.
 This must be the case if usable energy is not conserved at level $N+1$.
  \end{enumerate}

\noindent Section \ref{sec:linear poss} considers when higher level
linearity can emerge from lower level linear equations (Cases 1 and
2), while Section \ref{sec:open} makes the case that generically we
may expect higher level behaviour to \emph{not} be Hamiltonian (Case
3). Section \ref{sec:not linear} considers examples where this does
not occur; this vindicates Point 5 in Section \ref{main1}.

\subsection{Cases where linearity can emerge}\label{sec:linear poss}
I now consider a series of cases where linear higher level behaviour emerges from linear lower level behaviour.\\

\textbf{Classical to classical example: Centre of Mass motion.}
The classical example of emergence of higher level linear
behaviour out of lower level linear behaviour is the case of
centre of mass motion (see \cite{Gla60} for a clear description).
Consider a system of $N$ point particles of mass $m_i$ at position
$\textbf{r}_i$. Newton's law of motion for the \emph{i}th particle
is
\begin{equation}
m_i \ddot{\textbf{r}}_i = \textbf{F}^\ast_i = \textbf{F}_i + \sum_j \textbf{F}_{ij} \label{N1}
\end{equation}
Here $\textbf{F}^\ast_i$ is the total force on the \emph{i}th particle, $\textbf{F}_i$ is the external force,
and $\textbf{F}_{ij}$ is the internal force due to the \emph{j}th particle (there is no self-force:
$\textbf{F}_{ii} = 0$).
Newton's third law states action and reaction are equal and opposite:
\begin{equation}
\textbf{F}_{ij} = - \textbf{F}_{ji}. \label{N2}
\end{equation}
Consequently adding the equations (\ref{N1}) together for $i = 1$ to $N$,
\begin{equation}
\sum_i m_i \ddot{\textbf{r}}_i = \sum_i  \textbf{F}^\ast_i = \sum_i \textbf{F}_i  \label{N3}
\end{equation}
Defining the total mass $m$, centre of mass position $\overline{\textbf{r}}$, and total external force $\textbf{F}$ by
\begin{equation}
m := \sum_i m_i, \,\,  m \,\dot{\overline{\textbf{r}}} := \sum_i  M_i \dot{\textbf{r}}_i,\,\,
\textbf{F} := \sum_i  \textbf{F}_i  \label{N4}
\end{equation}
we find
\begin{equation}
 m \,\ddot{\overline{\textbf{r}}} = \textbf{F}  \label{N5}
\end{equation}
so the linear law for the individual particles (first equality in (\ref{N1}))
is replicated by the coarse-grained variables (\ref{N4}) - irrespective of the
nature of the internal forces.\\

This leads to the emergence of Hamiltonian dynamics for particle
motion \cite{Fowles,Cal96}, as for example applied in celestial
dynamics (governing the dynamics of stars in galaxies
\cite{BinTre87}). This also  applies
to motion of objects on Earth in situations where friction may be ignored: but they are very limited. \\

\textbf{Classical to classical example: Geometric optics.} In the
high frequency limit, Maxwell's equations for the electromagnetic
field leads to geometric optics \cite{Jac67,LipLip69,Hec75}, with
light propagating in a way described by Hamiltonian dynamics. The
different wavelengths do not interfere with each other because the
system is linear,
hence spectral decomposition makes sense.\\

\textbf{Classical to classical example: Engineering and Natural
systems.} As pointed out strongly by Bracewell \cite{Bra86}, many
manufactured and engineering systems have a linear dynamics that
leads to periodic behaviour and the suitability of Fourier
Analysis. This occurs particularly when the system is engineered
to have linear modes, for example organ pipes, guitars, linear
electrical and electronic circuits, and so on; however there may
be such modes in other cases, for example wave modes in suspension
bridges and torsional oscillations of buildings. There are also
similar instances in the natural world, for example propagation of
water waves and sound waves - indeed anywhere where Fourier
Analysis applies, linearity of the relevant degrees of freedom
leading to the splitting
of the system into normal modes with different frequencies that don't interfere with each other. \\

However these examples although ubiquitous are also limited: the
engineering examples are carefully tailored to behave in this way,
often
at considerable expense, and they have frequency limits beyond which the linear behaviour ceases. Similarly the linear behaviour of natural systems is very limited in time and space. Non-linearities intrude when we examine behaviour beyond these limits.\\

\textbf{Quantum to classical example: Ehrenfest's theorem.}
As a consequence of the Schr\"{o}dinger equation
(\ref{evolution}), the time derivative of the expectation value for a quantum mechanical operator is
determined by the commutator of the operator with the Hamiltonian of the system:\footnote{For a
conveniently accessible proof, see the Wikipedia entry on Ehrenfest's theorem.}
\begin{equation}
\frac{d}{dt} \langle A\rangle = \frac{1}{i\hbar} \langle [A,H] \rangle
+ \langle\, \frac{\partial A }{\partial t}\,\rangle \label{E1}
\end{equation}
Applying this to the case of a particle of mass $m$ and momentum $p$ moving in a
potential $V$ (see (\ref{L})) so that $H = p^2/2m + V$, and defining
$\langle F \rangle = - \langle \nabla V \rangle$, one finds
\begin{equation}
\frac{d  \langle p \rangle }{dt}  = \langle F \rangle, \,\,
\frac{d^2\langle x \rangle}{dt^2}  = \frac{1}{m} \langle F \rangle,
 \label{E2}
\end{equation}
 in agreement with the classical equation (\ref{N5}). Hence the linearity
of (\ref{evolution}) results in the linearity of the relations (\ref{E2}),
which are however not quantum relations (they have a classical form).\\

\textbf{Quantum to quantum: Renormalization group} In some cases one
can prove that coarse-graining a Hamiltonian systems leads to
another Hamiltonian system with the same Hamiltonian but different
values of its constants. One example is the Wilson approach to
renormalization theory, where the high momentum degrees of freedom
in the generating functional $Z[J]$ are integrated out, leading to
the renormalization group relating parameters of the original
Lagrangian to the new Lagrangian (\cite{PesSch95}:394-409; \cite{Zee03}:341-345).
However this is possible only in restricted circumstances (\cite{PesSch95}:402-403).\\

Another example is the Kadanoff construction, explicitly coarse
graining an Ising model, thus defining a coarse-grained lattice and
block spin variables. The coarse-grained dynamics are governed by a
Hamiltonian that is a function of the coarse grained variables on
the coarse grained lattice (\cite{ChaLub00}:237-242); indeed the
block spins interact via the same Hamiltonian as the original spins,
leading to a scaling of free energy and applicability of the Wilson
renormalization group \cite{Wil75}, see (\cite{ChaLub00}:245-248). \\

\textbf{Quantum to quantum: Effective Theories}
In some cases, coarse graining will result in a Hamiltonian theory at
the higher level, but with a Hamiltonian that has a different form. This is the case of
\emph{effective field theories} that emerge at higher level from the
underlying physics (\cite{Har01}, \cite{Zee03}:437-440):
an effective Lagrangian or Hamiltonian governs the higher level dynamics,
but it's different from the one you started with.
One cannot always derive this higher level effective action by explicit
coarse graining, but can often determine the form the effective
action should take by symmetry and conservation principles.
The classic example (\cite{Zee03}:441) is Fermi's $\beta$-decay theory \cite{Wil68},
now embodied in \emph{Fermi's Golden} Rule (\cite{Sak94}:332), which is of
wide application (see e.g. \cite{CraThi84}:84-86; \cite{GemMicMah04}:20,165-166).\\

Other examples are effective field theories of a Hall fluid (\cite{Zee03}:302-303) and of proton decay
(\cite{Zee03}:440-441). A more recent application relates to gravitational theory and the early universe.
When one treats cosmological inflation in the early universe as being due to an effective theory,
integrating out physics above some energy scale
$\Lambda$ induces nonrenormalizable operators in the effective theory. This can also lead to
corrections to the kinetic terms which contain higher powers of derivatives; the effects
on the early universe are different than in the standard theory (\cite{Fra09,Fra10}
and references therein).\\

\textbf{Quantum to quantum: Long range order} The electron system in
superconductors can exhibit \emph{long range order}, with strong
correlations in the wave functions of pairs of particles over
distances longer than the coherence length (\cite{Zim79}:402-403).
Hence one can introduce a macroscopic wave function
$\Psi(r)$ (the Ginzburg-Landau order parameter) for the superfluid
component of the electron density, leading to flux quantization
(\cite{Zim79}:404-405) as a macroscopic manifestation of
quantum mechanics. $\Psi(r)$ obeys a time dependent Schr\"{o}dinger
equation ((11.87) in \cite{Zim79}) which underlies the Josephson
effect (\cite{Zim79}: 405-410).\\

This is possible only in the context of metals with a periodic
lattice structure, or other materials
that allow superconductivity (\cite{Zim79}:396,410-414). The
restricted nature of the contexts that allow this emergence of
higher level effective quantum equations is shown in the great
difficulty of the search for superconductors other than metals. In
the case of metals, it is only possible when the temperature is
exceedingly low, so that the non-linear interactions that would
occur at higher temperatures are suppressed.

\subsection{Hamiltonian environment?}\label{sec:open}
The focus in this section is to make the case that generically the
environment of a quantum system may be expected to not behave in a
Hamiltonian way.

\subsubsection{Example of Non-Hamiltonian emergence} An example of the latter type is as follows:
Two systems $A$ and $B$ with respective states $\ket{\ps_A}$ and
$\ket{\ps_B}$ in Hilbert spaces $\cal{H}_A$ and $\cal{H}_B$ have a
joint wave function
\begin{equation}
 \ket{\ps_{AB}} = \ket{\ps_A} \otimes
\ket{\ps_B} \label{product}
\end{equation}
 A general state is
\begin{equation}
 \ket{\ps_{AB}} = \sum_{i,j} c_{nm} \ket{u_n(x)}_A \otimes
\ket{u_m(x)}_B \label{entangle}
\end{equation}
 Two states are entangled when
their wavefunctions cannot be written as a simple product state
(\ref{product}). If their evolution is given by
\begin{equation}
 \ket{\ps_2}_A =
U(t_2,t_1)_A\ket{\ps_1}_A,\,\, \ket{\ps_2}_B =
U(t_2,t_1)_B\ket{\ps_1}_B,\,\,
\end{equation}
then \begin{equation}
 \ket{\ps_2}_{AB} = U(t_2,t_1)_{AB} \ket{\ps_1}_{AB}, \,\,\,
U(t_2,t_1)_{AB} := U(t_2,t_1)_{A}\otimes U(t_2,t_1)_{B}
\label{joint}
\end{equation}
 so the joint evolution is unitary. However if a
wave function projection (\ref{trans}) takes place for either
component then it's evolution is not unitary and neither is that of
the composite state: it cannot be represented as (\ref{joint}). Thus
non-unitary evolution emerges at the higher level from the
non-unitary evolution at the lower level (which is reflected in the
way the density matrix evolves through a Markovian master equation
in Lindblad form (\cite{GemMicMah04}:54-58; \cite{BrePet06}:297-299; \cite{WisMil10}:119-121),
with consequent entropy generation \cite{Zeh90}.\\

This example is perhaps controversial because it involves the
disputed nature of the quantum measurement process. I will give
other examples of non-linear emergence in the next section and
Section \ref{sec:not linear}, and then pick the theme up in Section
\ref{sec:classical emerge}.

\subsubsection{Open systems and their environment}\label{sec:open1}
\textbf{Effect of the environment on the system}:
Following Breuer and Petruccione, consider an open quantum system
${\cal S}$ (`the system') coupled to another quantum system  ${\cal B}$ (`the
environment'), with respective Hilbert spaces
${\cal H}_{\cal S}$ and ${\cal H}_{\cal B}$ (\cite{BrePet06}:110-120).
The Hilbert space ${\cal H}$ of the combined system
${\cal T} = {\cal S} + {\cal B}$ is ${\cal H} = {\cal H}_{\cal S} \otimes {\cal H}_{\cal B}$.
The total Hamiltonian $H_{\cal T}$ is taken to be of the form
\begin{equation}
 H_{\cal T} = H_{\cal S} \otimes I_{\cal B}+ I_{\cal S} \otimes H_{\cal B} + \hat{H}_I(t)\label{product1}
\end{equation}
where $H_{\cal S}$ is the self Hamiltonian of the open system, $H_{\cal B}$ the free Hamiltonian of
the environment, and $\hat{H}_I(t)$ the Hamiltonian describing the interaction between the system
and the environment. Now an ensemble ${\cal E}$ of pure ensembles ${\cal E_\alpha}$ for the total system
${\cal S}$ with weights $w_\alpha$ has a density matrix
\begin{equation}
\rho = \sum_\alpha w_\alpha \ket{\psi_\alpha} \bra{\psi_\alpha}. \label{rho}
\end{equation}
The reduced density matrix for the system ${\cal S}$, given by tracing out the environment, is
\begin{equation}
\rho_{\cal S} = tr_{\cal S} \rho \label{rhoS}
\end{equation}
It follows from the unitary evolution of the total density matrix $\rho$ that the reduced
density matrix evolves according to the Lindblad master equation
\begin{equation}
\frac{d}{dt}\,\rho_{\cal S}(t) = -i \left[H, \rho_{\cal S}(t)\right]
\, + \, {\cal D}(\rho_{\cal S}(t)) \label{drhoS}
\end{equation}
where the unitary part of the dynamics is generated by the new Hamiltonian $H$ and the
dissipator ${\cal D}(\rho_{\cal S})$ is determined by the spectral decomposition of the
density matrix $\rho_{\cal B}$ of the environment (\cite{BrePet06}:103-119). The
viewpoint here is that shown in Diagram 3.\\

The two key points then are that (i) in general $H \neq H_{\cal
S}$ -- this is what opens the way to the renormalization group and
higher level effective Hamiltonian theories -- and (ii)
generically ${\cal D}(\rho_{\cal S}) \neq 0$: the higher level
system is not Hamiltonian, and hence (\ref{drhoS}) is associated
with the generation of entropy
(\cite{BrePet06}:123-125;\cite{Zeh90}). This carries through to
all the other versions of the master equation, for example the
interaction picture master equation (\cite{BrePet06}:130) and the
quantum optical master equation
(\cite{BrePet06}:140-149).\\

\begin{center}
\begin{tabular}{|c|c|c|}\hline
  (Hamiltonian)  & System plus environment ${\cal T}$ &  \\ \hline
     &    $\Downarrow$  $\Downarrow$& (Coarse grained)     \\ \hline
  System ${\cal S}$  &  $\Longleftrightarrow$ components $\Longleftrightarrow$ &    Environment ${\cal B}$ \\ \hline
  (Non-Hamiltonian) &   &  \\ \hline
\end{tabular}\\
\end{center}

{}\\ \textbf{Diagram 3:} \emph{The system plus environment evolve in
a Hamiltonian way, and interact with each other. When the
environment is
traced over, the system evolves in a non-Hamiltonian way}.\\

Another example is the Hawking effect \cite{Haw75}, in which tracing
over modes (of a pure state quantum field) that are lost behind a
black hole horizon results in a thermal state in the external
region.\footnote{I thank the referee for this suggestion.}\\

\textbf{Effect of the system on the environment}: Now change
viewpoint: coarse grain the system not the environment. But the
same equations apply! Just swap the labelling ($B \leftrightarrow
S$) in the above equations and the result will be
\begin{equation}
\frac{d}{dt}\,\rho_{\cal B}(t) = -i \left[\widetilde{H}, \rho_{\cal B}(t)\right]
\, + \, {\cal \widetilde{D}}(\rho_{\cal B}(t)) \label{drhoS1}
\end{equation}
(where now the Hamiltonian $\widetilde{H}$ and dissipator ${\cal
\widetilde{D}}$ are different than in the previous case). This
shows the system can cause non-Hamiltonian behavior in the
environment. Realizing that in terms of the hierarchy in Table 1,
the environment is at a higher level than the system, we can
represent the situation
as in Diagram 4. \\

But this raises the key issue: \emph{why are we entitled to assume
the combined system ${\cal T}$ behaves in a Hamiltonian way}? We
could start with ${\cal B}$  as the total system and then separate
out a subsystem of ${\cal B}$ to be designated as the subsystem
${\cal S}_1$ of interest.  We are surely entitled to query why we
should assume that the top level (${\cal T}$ in Diagram 3, ${\cal
B}$  in Diagram 4) behaves
in a Hamiltonian way?\\

\begin{center}
\begin{tabular}{|c|c|c|}\hline
    & Environment ${\cal B}$ & (Non-Hamiltonian) \\ \hline
    (Coarse grained)  &   $\Uparrow$  &      \\ \hline
  &  System ${\cal S}$   & (Hamiltonian)   \\ \hline
\end{tabular}\\
\end{center}

{}\\ \textbf{Diagram 4:} \emph{When coarse-grained, the Hamiltonian
system {\cal S} induces non-Hamiltonian behaviour in the environment
${\cal B}$}.

\subsubsection{Emergence of non-linearity} Now this contradicts what
is usually understood as the standard hypothesis of quantum physics:
as stated by the referee of this paper, this is as follows:

\begin{quote}
\textbf{Standard Hypothesis}: \emph{In a closed system obeying
quantum mechanics at the lower level, nonlinearity cannot emerge at
a higher level from quantum mechanics plus interactions alone. Only
if the system is open can nonlinearity feed into the system. A full
quantum description of an isolated macroscopic system may be
practically impossible, but either it is possible in principle, or
quantum mechanics breaks down at some level of size or complexity.
In an isolated system one cannot appeal to openness to inject the
crucial nonlinearity. Where do the the nonlinearities come from, if
they are not present in the constituent particles and their
interactions alone}?
\end{quote}
But a main point of this paper is that the assumption of linearity
at all scales is an \emph{a priori} untested assumption that
extrapolates what happens at micro scales to arbitrarily large
scales, and may or may not be true. It is an assumption involving
extrapolation of extraordinary scope when applied to the universe as
a whole. The standard view is that things are linear on the largest
scales, and non-linearity sometimes emerges by top-down action from
these large scales to smaller scales. The view in this paper is the
converse:
\begin{quote}
\emph{\textbf{Alternative view}: Linearity holds on the smallest
scales, and higher level behaviour emerges from the local
applicability of such linear behaviour everywhere; this higher level
behaviour may or may not be linear. The non-linearity arises from
specific configurations of particles that lead to complex networks
of interactions; these configurations are higher level properties of
the system. It is initially an experimental question whether higher
level behaviours are linear or not. Theory must then accommodate to
whatever experiments determine; and they appear to show that quantum
mechanics does indeed break down at higher levels of size or
complexity.}
\end{quote}

\noindent An example is the non-linear explosive behaviour of a
mixture of trinitrotoluene (TNT) and oxygen in a closed container,
which is not due to the mixture being an open system (it is not), it
is due to the molecular structure of the TNT. In section
\ref{emergeh} and in the next section I give examples where the
usual assumption is indeed wrong. The situation shown in Diagram 3,
where the higher level is Hamiltonian, is not generic; it will only
hold under special circumstances. The suggestion is that this
vindicates the claim made in Point 4, Section \ref{main1}. One view
would be that the examples in the next Section are merely
phenomenological practical procedures leading to effective theories
for dealing with large systems, as opposed to fundamental. The idea
here is that they may indeed be fundamental: each level should be
regarded as having an existence in its own right, rather than merely
derivative from the level below; emergence is real emergence, rather
than an illusion \cite{And72,Lau00,Ell04}. The TNT has real causal
power not implied by the constituent particles and their
interactions alone: it is in the \emph{organization} of those
particles that the crucial causal power resides (the same
constituent particles are still there after the explosion, subject
to the same interactions; it is their organization that is
different). This possibility of real emergent higher level causal
powers occurs because top-down causation takes place in the
hierarchy of complexity \cite{EllNobOCo12}, as discussed below
(Section \ref{sec:contextual}).

What then determines how nonlinearity emerges from linearity? It
resides in the details of the physics of emergence \cite{Sco07},
which leads to new levels of complexity with their own logic of
behaviour and associated causal powers. The details are very
different in different cases, as the examples that follow show.

\subsection{Cases where unitary behaviour does not emerge}\label{sec:not linear}
Here I consider some examples where the coarse grained behaviour
arising from the underlying theory is not Hamiltonian and so does
not exhibit quantum characteristics. There will be four levels at
which this happens.
\begin{itemize}
\item The Anderson idea of novel notions at higher levels
\cite{And72} (e.g. Section \ref{emergeh});
\item The non-interference of states at higher levels (e.g. Section
\ref{class1});
\item Non-unitary evolution at higher levels (e.g. Section
\ref{class2});
\item The general idea of classical emergence at higher levels (Section \ref{sec:classical
emerge}).
\end{itemize}
Much of
 this is perhaps rather obvious; it is worth pursuing for two reasons. First, as I will
 consider later (Section \ref{sec:classical emerge}),
 any situation where unitary behaviour does not emerge is a possible channel for creating
 classical systems out of quantum components. Second, some of the present day literature (e.g. \cite{Wal01,Har08})
 assumes that quantum behavior will be always be present at higher levels. The present view contradicts
 that assumption (see Section \ref{sec:class applications}).

\subsubsection{Stochastic situations: equilibrium}\label{class1}
\textbf{Boson Gas}: In the case of a boson gas, the wave function at the quantum level is symmetric,
(\cite{Dir58}:205-211), resulting in the Bose-Einstein distribution law (\cite{AloFin71}:528-530) on
coarse-graining. Non-linear macroscopic laws of behaviour emerge, describable in purely classical terms.
For example, in the case of photons one obtains the black body spectrum for radiation
(\cite{AloFin71}:7-11;531-532), and the consequent formula for energy density and pressure of a photon gas:
\begin{equation}
\rho(T)  = \frac{8\pi h}{c^3}\int_0^\infty \frac{\nu^3 d\nu}{e^{h\nu/kT}-1}, \,\,\,\,p(T) = \frac{\rho(T)}{3c^2}
\label{bbr}
\end{equation}
The key point is that \emph{these are relations for classical variables}: there is nothing in the behaviour
at this higher level corresponding to superposition of states or entanglement.
The pressure $p(T)$ and density $\rho(T)$ given by (\ref{bbr})
 are not in any sense fictitious variables. Rather they are the essential causally effective variables at their
level of description in the hierarchy (Table 1), for example playing
a key role in astrophysics and cosmology \cite{Dod03,EllMaaMac11}.
The situation is shown in Diagram 5:\\

\begin{center}
\begin{tabular}{|l|c|c|}\hline
   Classical Level : & Gas laws & \emph{Temperature $T$, Density $\rho$, Pressure $p$}\\ \hline
     &\emph{\textbf{Coarse Grain}} $\Uparrow $   & $\Uparrow$ Bose Einstein statistics\\
  \hline Quantum Level :  & Photon gas &   \emph{symmetric wave function}  \\ \hline
\end{tabular}\\
\end{center}

{}\\ \textbf{Diagram 5:} \emph{The emergence of higher level
effective classical
variables from the underlying quantum theory}.\\

Similarly one attains  macro formula for the pressure and density of
a gas of molecules with zero integral spin
(\cite{AloFin71}:Eqn.(13.32)). In a metal, a phonon gas leads to
formula for the heat capacity $C_V$ of a solid
(\cite{AloFin71}:Eqn.(13.28)). These are all emergent classical
properties, as in the case of the
energy density and pressure in (\ref{bbr}). \\

\textbf{Fermi Dirac gas}: In the case of an electron gas, the wave function at the quantum level is anti-symmetric
(\cite{Dir58}:205-211), resulting in Fermi-Dirac statistics
(\cite{AloFin71}:519-522). This again results in higher level non-linear behaviour describable in
purely classical terms, e.g. the
thermoelectric current density coming from a metal surface in
terms of the temperature of the metal (\cite{AloFin71}:Eqn.(13.11)).\\

Overall, the emergence of these classical levels from the underlying quantum theory is in accord with the
view put in Section \ref{sec:context}:
\begin{quote}
\emph{Each of the higher levels of the hierarchy of complexity is real in its own right, described
by relevant variables for that level, and laws of behaviour that are effective at that level. These
variables and interactions emerge from the underlying quantum variables, and in the case of equilibrium states
are classical variables}.
\end{quote}
But the word `` effective'' sounds perjorative: they are
\emph{the} laws of behaviour applicable at that level. When
equilibrium occurs, classical higher level thermodynamic behaviour
emerges from the underlying quantum structure. The way this
happens is presented by Gemmer, Michel and Mahler
\cite{GemMicMah04}. The essential point is that  the statistical
interactions between the components that lead to equilibrium
destroy any coherence among the higher level variables. An example
is that the transition to equilibrium in a crystal relies on the
Umklapp process (\cite{GemMicMah04}:223), which does not preserve
momentum, and so is not a unitary process. Presumably this
corresponds to frequent collapse of the wave function at the
micro-level: for if that does not take place, the necessary
interactions between the components for thermalization will not
have occurred, and they can be expected to occur very frequently.

\subsubsection{Dissipative effects: non-equilibrium}\label{class2}
Dissipative effects, and consequent entropy increase enshrined in
the second law of thermodynamics (\cite{Fuc96}:139-144,
\cite{Atk94}:119-124) are a fundamental part of what goes on at the
macroscale \cite{Edd28}. This is associated with coarse graining of
microphysics \cite{Pen04}: 688-699; \cite{GemMicMah04}:45-48),
whereby energy stored in microscopic states are inaccessible via
macro variables, and so lead to unusable energy and hence effective
energy loss experienced at the macro scale (even though no energy is
actually lost if one takes into account also the inaccessible
internal energy degrees of freedom). These states thus act like the
environment in Section \ref{sec:open1}. The process whereby this
emerges from the underlying
unitary theory is coarse graining by integrating out micro variables.\\

Clearly this is a non-Hamiltonian macrolevel behaviour, emerging
from Hamiltonian microlevel behaviour. It derives from the
underlying unitary quantum systems, and a large literature on
dissipative quantum systems discusses how this happens (see
\cite{BrePet06}:166-194,465-480); but as in the previous case
(Section \ref{class1} and Diagram 5), it results in effective
classical behaviour at the higher level (this has to be so, as that
behaviour is dissipative and hence non-unitary; clearly
superposition, interference, and entanglement cannot be expected to
occur in terms of the interaction dynamics of these macro
variables). It results for example in the existence of dissipative
systems in biology (\cite{Pea89}:40-56,62-72), which are essential
for life to exist (\cite{Pea89},\cite{CamRee05}).

\subsubsection{Chaotic systems}\label{chaos}

Quantum systems do not exhibit chaotic behaviour (subject to
exponential sensitivity to initial conditions), but non-linear
classical systems can do so. Quantum chaos theory examines the
problem of how this can be possible, in view of the correspondence
principle relating classical to quantum mechanics \cite{Haa01}.

\subsubsection{Elements or circuits with a threshold}\label{class3}
Whenever an element or circuit has a threshold, the behaviour is
non-linear when one crosses the threshold. Examples are rectifiers
and digital elements such as inverters, logic gates, and more
complex digital circuits (\cite{Mel90}:34-57). Important examples
are photo diodes and photovoltaic cells (\cite{Mel90}:200), based in
the photoelectric effect (\cite{AloFin71}:11-14). Other examples are
chemical and nuclear systems with an activation threshold.

\subsubsection{Feedback control loops}
A key example is feedback control systems with fixed goals, such as
a thermostat. Such systems are classically described, and hence
non-unitary; they will also usually be dissipative, so there will be
no question of a Hamiltonian description. They are ubiquitous in
engineering \cite{HarBol69} and in biology
\cite{Mil66,Cal76,IngIng10}, and are an important form of top-down
action \cite{Ell08}.
Their functioning is indicated in Diagram 6.\\

\begin{center}
\begin{tabular}{|c|c|c|}\hline
   Controller  & $\Leftarrow$  \emph{Correction signal} & \\ \hline
    \emph{\textbf{Action}}  $\Downarrow$ & \emph{\textbf{Feedback}}  $\Uparrow $  &   \\ \hline
   \emph{State} & $\Leftrightarrow$ Comparator $\Leftrightarrow$ &   \emph{Goal} \\ \hline
\end{tabular}\\
\end{center}

{}\\ \textbf{Diagram 6:} \emph{The basic features of a feedback
control system. The goals tend to lead to a specific final state via
a specific mode of physical action. The initial state of the system
is then irrelevant to its final
outcome, provided the system parameters are not exceeded}.\\

At each cycle of the system, a measurement of the system state is
compared with a desired state, and an error signal sent to a
controller to correct the error and make the system state approach
the desired state \cite{BelKal64,DisStuWil90}. Hence the feedback
control process demands a determination of the current state of the
system, to give the information contained in the feedback control
signal to the controller. This specific information utilized to
determine the further dynamics can only be obtained by a measurement
process entailing collapse of the wave function  at the underlying
quantum level. Non-unitary collapse interposes at each time step. Furthermore the
macro dynamics clearly will not support superposition or
constructive interference: whatever the input state, the output
state is the same (the desired temperature, in the case of a
thermostat). This outcome is due to the choice of goals - a high
level property of the system that is not reducible to lower level
entities, or even describable in lower level language \cite{Ell08}.
An example is the choice of setting of the desired temperature in a
thermostat, which can be chosen at will; this sets the goal (the
chosen temperature), which the system then
implements (many electrons flow to make this happen).\\

 What then about the burgeoning literature on quantum feedback
control (see \cite{WisMil10}:216-340)? Does this not contradict
what has just been said? No, it does not. Inspection will show
that all such schemes use classical detectors to feed back a
control signal to the quantum system (e.g. \cite{SarMil05};
\cite{WisMil10}: Fig 5.1, Fig 5.2, Fig 6.1; \cite{Sayetal11}).
This has to be the case, as a purely quantum system could not
provide the needed classical control signal. It can't do this
unless specific individual measurements take place to provide the
classical signal! And one should note the following point: suppose
one linearizes to the case of small disturbances about the
equilibrium state. It will still be true that a measurement is
needed to complete the circuit, so quantum theory won't be able to
handle it; and it will still be true that, because the dynamics
drives all input values to zero, there will be no superposition of
solutions. Hence even the linearized version of the equations will
not be of the unitary form (\ref{U1}).

\subsubsection{Complex networks}
A feedback control loop is just one of the network motifs identified
by Alon as occurring in biological networks \cite{Alo07}.
Real biological networks are immensely complex \cite{Dav06}, and will contain
many complex interactions and network motifs, including feedback control loops. Hence they too
will be non-Hamiltonian systems. This will apply in particular to the connections
between neurons in a brain \cite{KanSchJes00}, which are made up out of microcircuits
that are themselves very complex \cite{SheGri10}.

\subsubsection{Adaptive Selection}\label{adapt}
\begin{center}
\begin{tabular}{|c|c|c|}\hline
   \emph{System state } & $\Leftarrow$  \emph{\textbf{Selection agent:  selects state}} &  \\ \hline
    \emph{\textbf{Variation}}  $\Downarrow$ &   $\Uparrow $  &  Meta-goals: \\ \hline
   \emph{Ensemble of System States} & $\Rightarrow$ Preferred state in ensemble $\Leftarrow$ &
   \textbf{Selection criteria} \\ \hline
    & $\Uparrow $ &    \\ \hline
    & \textbf{Environment } &    \\ \hline
\end{tabular}\\
\end{center}

{}\\ \textbf{Diagram 7:} \emph{The basic features of adaptive
selection. Selection takes place from an ensemble of states, the
selection being based on the action of some selection
criteria in the context of the specific current environment}.\\

A further key example is the
process of adaptive selection \cite{Kau93,Gel94}, ubiquitous in biology \cite{CamRee05}, but also
occurring in digital computers, for example in artificial neural
networks and genetic algorithms \cite{Ell08}. Selection takes place from an
ensemble of initial states to produce a restricted set of final
states that satisfy some selection criterion. The process is summarized in Diagram 7.
Note that it can take place in a once-off form: in biology it gains its enormous strength because
it is repeated so many times, but that repetition is not essential to the concept of selection.
In a selection event, in effect the selection agent compares the
entities available in the initial ensemble to determine the best
candidates on the basis of the preset selection criteria, evaluated
in the current environmental context. The best candidate is selected
and retained as the outcome of the event; the rest are discarded.
The meta-goals embodied in the selection criteria do not necessarily
lead to a specific final state (although they may do in some
restricted circumstances): rather they lead to any one of a class of
states that tends to promote the meta-goals. Thus the final state is
not uniquely determined by the initial data; random variation
influences the outcome by leading to a suite of states from which an
adaptive selection is made in the context of both the selection
criteria and the environment \cite{Hol92}. \\

One could call it simply \emph{selection}, but I prefer
\emph{adaptive selection} to emphasize that it always take place
as a consequence of the existence of selection criteria, which are
higher level entities in the hierarchy of causation; hence this is
another form of top-down action \cite{Ell08}. An example is the
case of state vector preparation by a polarizer, which I will show
below (Section \ref{prep}) can be regarded as a case of adaptive
selection, because it selects the desired specific polarization
state from a jumble of incoming random polarization states. The
experimenter chooses the axes of the polarizer; this determines
which polarization state gets selected from those arriving. This
is a simple model of the general way in which adaptive selection
is guided by the meta-goals; in most cases they are not as
specific (in biology for example, it is simply survival). Note the
difference from feedback control, where no
ensemble of incoming states is involved.\\

Like the case of feedback control, this also demands an effective
collapse of the wave function, firstly as the selection process
results in specific determinate outcomes, and secondly because the
process in effect makes a decision on the basis of the specific
outcomes of the individual variations that underlie such selection
processes. Superposition of outcome states is hardly possible: it
is not a Hamiltonian process. As in the case of feedback control
loops, even if one linearizes there will be no Hamiltonian
description possible; inter alia this is because a selection
process involves thresholds, which are non-Hamiltonian (as discussed above
in Section \ref{class3}).\\

The importance of this process is that it is the way meaningful
information enters the physical world in a way that is unpredictable
on the basis of the underlying physics, thereby enabling the
emergence and functioning of true complexity \cite{CamRee05,Ell08}.
This takes place by selection of a subset of states from an
ensemble, \emph{which is the basic process whereby information that
is relevant in a specific context} \cite{Roe05} \emph{is selected
from a jumble of irrelevant information}. Some information is
selected, some discarded. This is what enables an apparent local
violation of the second law of thermodynamics, as in the case of
\emph{Maxwell's Demon} (\cite{FeyLeiSan63}:46-5;\cite{LefRex90},
\cite{AhaRoh05}:4-6; \cite{Car10}:186-189, 196-199) -- who is indeed
an example of an adaptive selection agent, acting against the local
stream of entropy growth by selecting high-energy molecules from
those with random velocities approaching a trap-door between two
compartments. The selection criterion is the threshold velocity
$v_c$ deciding if a molecule will be admitted into the other
partition or not. It is significant that Maxwell's demon type
devices can be created in the lab
\cite{RusMugRai06,Pri07,Pri08,Schetal11}, explicitly demonstrating
that adaptive selection can arise in a quantum physics context.  It
occurs also in microbiology, where active transport systems are
enabled by voltage gated ion channels  (\cite{Leh73}:191-206).
\\

\textbf{Darwinian selection}  is just the process of repeated
adaptive selection in biology , with reproduction
 and variation between each stage of selection \cite{CamRee05}; it certainly takes place in the real world as
 an emergent feature from the underlying quantum Hamiltonian dynamics, and is the core feature
 leading to the existence of life. It is obviously not a Hamiltonian process.

\section{Contextual effects in quantum physics}\label{sec:contextual}
Now I turn to the converse of emergence, namely  the way that
contextual effects change the nature of interactions at the lower
levels. Section \ref{topdown} considers the broad nature of top-down
causation in general, and Section \ref{sec:QM TD} specific cases
where it occurs in quantum physics. This relates to some of the
examples given in the previous section.

\subsection{Top-down causation}\label{topdown}
The higher levels of the hierarchy of complexity and causation
(Table 1) provide the context within which the lower level actions
take place. By setting the context in terms of initial conditions,
boundary conditions, and structural relations, the higher levels
determine the way the lower level actions occur.\\

 A simple example is
a digital computer \cite{Tan90}: the lower level transistors and
integral circuits function in exactly the same way whatever higher
level program is loaded; but the higher level program determines the
outcomes - music, pictures, graphs, or whatever. A physics example
is the way that cosmological-level coarse grained variables control
nuclear reaction rates in the early universe by determining how the
temperature $T$ varies with time, thereby determining the way
cosmological nucleosynthesis pans out \cite{Dod03,EllMaaMac11}. \\

\noindent The general picture is that in Diagram 8:

\begin{quote}
\emph{The lower levels do the work, but the higher levels decide
what is to be done}.
\end{quote}

\begin{center}
\begin{tabular}{|l|l|c|}\hline
   Level N+1: & Higher level theory  &  \textbf{\emph{Effective Theory}} \\ \hline
     & \emph{Top-down effects }&  $\Downarrow$\\ \hline
  Level N:  & Quantum Theory  &\emph{\textbf{Contextual effects}} \\ \hline
\end{tabular}
\end{center} {}\\ \textbf{Diagram 8:} \emph{The effective higher level
theory exerts contextual effects
 on the operation of the underlying quantum theory}.\\

\noindent This can be regarded as top-down causation in the
hierarchy of complexity. Such causation, in conjunction with
bottom-up action, is the key to emergence of complexity from
underlying physics (for a full discussion and many examples, see
\cite{Ell08,EllNobOCo11}). The fundamental importance of top-down
causation is that it changes the causal relation between upper and
lower levels in the hierarchy, in particular enabling inter-level
feedback loops. It is a common view that ``\emph{only if the system
is open can nonlinearity feed into the system.}'';\footnote{This
comment comes from the referee.} but then the fundamental point is
that things are interconnected:
\begin{quote}
\textbf{Interacting systems:} \emph{there are no closed systems in
the real universe, apart from the universe itself.}
\end{quote}
 All finite systems are open because
their environment influences them both in historical terms, setting
the initial condition for the system to exist, and in functional
terms, affecting them on an ongoing basis, as acknowledged for
example in the discussions of environmental decoherence
\cite{Zur03,Zur04,GreZaj06}. This is a top-down influence from the
environment to the system. Furthermore the whole point of causal
networks, such as feedback control loops \cite{DisStuWil90} and
other network motifs \cite{Alo07}, is that they ensure that the
individual components are \emph{not} closed
systems: they feed information to each other. \\

\textbf{Proving top-down causation} How do we prove top-down effects
are occurring? One has to show that changing some higher level
condition changes lower level dynamics or behaviour. For example,
changing the length of an organ pipe changes the wavelengths of
possible standing waves, so the sound it emits depends on it size;
similarly changing the shape of a drum changes the sounds it emits.
By contrast, the black body spectrum (\ref{bbr}) is independent of
the size and shape of an oven that emits blackbody
radiation; it is determined by purely local effects.\\

\textbf{Equivalence classes} Technically, the way this works is that
equivalence classes of lower level states correspond to a single
higher level state \cite{AulEllJae08}; for example in the case of a
gas in a cylinder, a myriad of lower level molecular states $s_i$
will correspond to a specific higher level state ${\cal S}$
characterized by a temperature $T$, volume $V$, and pressure $p$,
which are the effective macroscopic variables. The number of such
lower level states that correspond to the higher level state
determines the entropy of that state. One can only access the
equivalence class by manipulating higher level variables rather than
the detailed lower level variables, hence cannot by higher level
action determine which specific lower level state $s_i$ realizes the
higher level state ${\cal S}$ (a proviso: one can design the kind of
apparatus that occurs in a quantum optics laboratory so that some
higher level
variables access specific lower level states; but these are exceptional situations).
Philosophers characterise this existence of equivalence classes through the phrase
``multiple realization''.\\

\textbf{Changing the basic elements} One further point of importance
is that it is not necessarily the case that one always has
unchanging lower level elements being combined in different ways to
form higher level complex structures. It may occur that the higher
level context actually changes the very nature of the lower level
entities that are combined to make the whole. An example from
physics is that a free neutron has completely different behaviour
than one bound in a nucleus: the former decays with a half life of
11 minutes, the latter last billions of years, hence it's essential
nature is changed by context. 
A chemistry example is that a free hydrogen is quite different than
a hydrogen atom incorporated in a water molecule. It is an
essentially different entity. 
In biology, this effect is of crucial importance: for example
initially identical cells are adapted to be different cell types
according to their position in the human body \cite{CamRee05}.

\subsection{Quantum mechanics examples}\label{sec:QM TD}
I now give a series of examples where contextuality in the sense
outlined above plays a role in quantum theory. I call this
\emph{top-down causation}, to distinguish it from the way the term
``contextuality'' is currently being used in many papers on quantum
theory (see \cite{Zei99,Kiretal09} and references therein). They are
undoubtedly related, but I wish to specifically refer to the kinds
of effect referred to in Section \ref{topdown} and in \cite{Ell08}.

\subsubsection{Particle-Wave duality}
Whether an entity acts as a particle or a wave is context dependent: this is the heart
of particle-wave duality, where one can determine whether particles going through a slit
should behave as particles or waves by the way one carries out the experiment (\cite{FeyLeiSan65}:1-1 to 1-7).
This has now been realised experimentally in the case of
a version of Wheeler's delayed choice experiment \cite{Whe78}
where the which-way choice is made after the particle has passed the slits \cite{Jacetal07}: a
case of top down causation from the apparatus to the very nature of the particle/wave at the
time it passed through the slits.

\subsubsection{Potentials emerging from forces}\label{effective}
One way top-down causation takes place is via the representation of the
interactions between many atoms in terms of an effective potential, treated as a classical entity.
Gemmer \emph{et al} give an illuminating example (\cite{GemMicMah04}:74-77)
in discussing the example of an ideal gas in a container.\\

The container provides the environment for the gas, and is made up
of an interacting set of particles (\emph{Fig.7.2} in
\cite{GemMicMah04}). Starting with a standard interaction
Hamiltonian, coarse graining leads to an effective ``box'' potential
$\hat{V}^g$ for each gas particle, comprising the mean effect of all
the container walls. This mean potential is then the higher level
context within which the gas particle moves; it can be represented
(\emph{Fig.7.3} in \cite{GemMicMah04}) by a smooth set of
equipotential lines, the transition from \emph{Fig.7.2} to
\emph{Fig.7.3} being a classic illustration of the coarse graining
process. One can regard the result as top-down action by the
potential (regarded as an entity in its own right) on the gas
particles. The underlying equivalence classes are all the different
configurations of particles that lead to the same effective
potential; it is these equivalence classes that are the significant
causal entity, rather than any detailed particle configuration that
leads to the potential. \\

Similar examples are the potential wells used in \emph{nuclear shell
models} (\cite{Dur00}:140-144), and the \emph{Slater treatment of
complex atoms}, explained by Pauling and Wilson thus
(\cite{PauWil63}:230):
\begin{quote}
    ``All of the methods we shall consider are based on the
    approximation in which the interaction of the electrons with each other
    has either been omitted or been replaced by a centrally symmetric force
    field approximately representing the average effect of all the other electrons
    on the one under consideration''.
\end{quote}
A similar method in astronomy is the
way a coarse-grained potential energy is derived for a galaxy, and then used to find
the motions of stars (\cite{BinTre87}:67-90,103-186;\cite{Sas87}:3-6).\\

These are examples of the method of \emph{mean field theory}
(\cite{ChaLub00}:198-208), which can be applied in many other
contexts. It can for example represent the way that electrical
wiring channels currents in electric circuits, through an
extremely complicated effective potential: an emergent higher
level entity. Indeed it enables one to represent arbitrary higher
level structures emerging from the underlying physical levels, and
then acting down on the lower level components by channelling the
way they interact with each other. One does not need to include a
representation of each individual interacting molecule. Examples
range from integrated circuits to  split-gate devices used in
nanotechnology  (\cite{Mil97}:96,104,112) to telephone systems,
chemical plants, and neuronal connections
via dendrites and axons in a brain.\\

Another example is the \emph{Caldeira-Leggett model}, a system plus
heat reservoir model for the description of dissipation phenomena in
solid state physics (\cite{BrePet06}:166-172, \cite{Cal10}). Here
the Lagrangian of the composite system $T$ consisting of the system
$S$ of interest and a heat reservoir $B$ takes the form
\begin{equation}
L_T = L_S + L_B + L_I + L_{CT},
\end{equation}
where $L_S$ is the Lagrangian for the system of interest, $L_B$ that for the reservoir (a set of non-interacting
harmonic oscillators), and $L_I$ that for the interaction between them.
The last term $L_{CT}$ is a \emph{counter term},
introduced to cancel an extra harmonic contribution that would come from the coupling to the environmental oscillators.
This term represents a top-down effect from the environment to the
system, because $L_I$ completely represents the lower-level
interactions between the system and the environment. $L_{CT}$ would
not be there if there was no heat bath; the effect of the heat bath
is more than the sum of its parts when $L_{CT}\neq 0$, because the
summed effect of the parts is given by $L_I$. Thus $L_{CT}$ should
be called the \emph{contextual term} rather than the counter term.

\subsubsection{Binding energies}\label{sec:bind}
When there are such extra terms in the interaction, this will
result in changes in energies. A crucial example is \emph{nuclear
binding energies}, the cost of putting emergent nuclear structures
together, which can be reclaimed on dismantling the structure.
These energies would not be there if the structure (a nucleus) was
not there, so it is a direct
result of the existence of the higher level structure, nucleons on their own have no such energy term. \\

Molecular binding energies are another example, of crucial importance in chemistry.

\subsubsection{Lattice waves and quasiparticles}\label{sec:quasi}
The periodic crystal structure in a metal leads (via Bloch's
theorem, (\cite{Zim79}:16-20) to lattice waves (\cite{Zim79}:27-75),
and an electronic band structure depending on the particular
solid involved (\cite{Zim79}:93-94,119-128), resulting in all
the associated phenomena resulting from the band structure. The
entire machinery for describing the lattice periodicity refers to a scale much larger
than that of the electron, and hence is not describable in terms appropriate to that scale.
Thus these effects all exist
because of the macro properties of the solid - the crystal structure - and hence
represent top-down causation from that structure to the electron states. \\

For example, this can lead to existence of quasiparticles such as
\emph{phonons} (\cite{Zim79}:59-62) that result from vibrations
of the lattice structure, and hence associated phenomena such as the
\emph{U-process} whereby momentum in electron scattering processes
is transferred to the system as a whole. It also leads to
\emph{Cooper pairs} produced by the exchange of phonons between
electrons (\cite{Zim79}:382-386) and hence to superconductivity
(\cite{Zim79}:386-394) and associated phenomena such as
superfluidity in metals (\cite{Zim79}:394-396). Because these are
all based in top-down action, they are \emph{emergent phenomena} in
the sense that they simply would not exist if the macro-structure
did not exist, and hence cannot be understood by a purely bottom-up
analysis, as emphasized strongly by Laughlin \cite{Lau00}.\\

Other examples are \emph{holes}, conduction electrons with negative
effective mass as determined by the energy surface ${\cal E}(k)$
(\cite{Zim79}:182-186), which are central to the physics of
semiconductors (\cite{Zim79}:59-62), and \emph{plasmons}
(particles derived from plasma oscillations).
The quantum Hall effect is a result of the existence of composite
Fermions, realised in the interface between two semiconductors
\cite{Jai00}. In all cases, it is the higher level context that
leads to their existence, because it determines the form of ${\cal E}(k)$. This represents
the effective result of the existence of the macro structure, similarly to the way effective potentials do
(Section \ref{effective}) .

\subsubsection{Decoherence}
Decoherence is the process whereby the environment (a macro context)
decoheres the wave function and selects preferred pointer states,
thus crucially determining the nature of micro outcomes
(\cite{Ish95} 155; \cite{BrePet06}, 212-270;
\cite{WisMil10}, 121-141). \\

Zurek argues this can be seen as a Darwinian like process he calls
environmental selection (\emph{Einselection}) \cite{Zur03,Zur04}.
This can therefore be seen as a case of top-down causation by
adaptive selection (Section \ref{adapt}): the lower level dynamics
does not by itself determine the outcome, which is shaped by the
higher level context of the environment.

\subsubsection{State Preparation}\label{prep}
State preparation in QM is a non-unitary process, because it can
produce particles in a specific eigenstate. Indeed it acts just like
state vector reduction (\ref{trans}), being a non-unitary transition
that maps a mixed state to a pure state. How can this happen in a
way compatible with quantum theory dynamics? \\

The crucial feature of quantum state preparation is pointed out by
Isham (\cite{Ish95}:74,134) as follows: selected states are drawn
from some collection ${\cal E}_i$ of initial states by some suitable
apparatus, for example to have some specific spin state, as in the
Stern-Gerlach experiment; the other states are discarded. This is
another case of adaptive selection, (see Section \ref{adapt}):
selection takes place from a (statistical) variety of initial states
according to some higher level selection criterion. As explained in
 Section \ref{adapt}, this is the characteristic way one can generate order out of a
disordered set of states by a process of selection from an ensemble
of systems, and so generate useful information \cite{Roe05}, just as
in the case of Maxwell's demon. This happens in two basic ways:
\emph{separation and selection}, which is unitary up to the moment
of selection, and \emph{selective absorption},
which absorbs energy and so is non-unitary all the time. \\

\textbf{Collimation, Deflection, and Selection} This is a very
general basis for state selection. In the case of the
\emph{Stern-Gerlach experiment} (\cite{FeyLeiSan65}:5-1 to 5-9),
collimation of an incoming stream of atoms by some slits is
followed by deflection in a non-uniform magnetic field, which
separates the initial beam into final beams according to their
spin; each final beam is then a polarized beam in a prepared spin
state. Thus when we choose to examine a particular spin by
selecting one of these beams, one set of incoming states is
selected and the other sets discarded. A \emph{mass spectrometer}
works on the same principle, separating out masses, as does a
\emph{spectrograph}, where a prism or diffraction grating sorts
out light by wavelength (so you can select a specific
pure colour by using a slit to collimate the light after it has passed through the prism).\\

Another example is a \textit{Nicol prism}, used to generate a beam
of polarized light (\cite{LipLip69}:132). A crystal of Iceland spar
is cut diagonally, the two parts being joined by Canada balsam. When
unpolarized light enters the crystal, it is split into two polarized
rays by \emph{birefringence}
(\cite{LipLip69}:131;\cite{Hec75}:111-118), the decomposition of a
light ray into two rays by an anisotropic crystal. The crystal is
shaped so that one beam is totally internally reflected and lost;
the other emerges parallel to the incidence direction. Birefringence
is caused by electromagnetic polarization in an anisotropic medium
with dielectric tensor $\epsilon_{ij}$ resulting from the
coarse-graining of the dipole contributions to the electric field
(Section \ref{emergeh}and \cite{Jac67}: 116-122)).\\

 Polarization is also caused by
\emph{reflection of light} at less than the critical angle at a
surface separating two transparent media. Then partial reflection
and partial transmission takes place
(\cite{LipLip69}:109-110;\cite{Hec75}:40-41,108-109), again
separating the initial beam into two polarized beams; so this can
also be used to prepare polarized states. The anisotropy in this
case is
caused by the layer separating the two media; the reflected light is polarized normal to the incidence plane.\\

\textbf{Selective absorption} Dichroism is the selective
absorbtion of one polarization state due to a linear structure in
a polarizer, which therefore selects a specific spin state from a
beam of incoming photons, thereby rejecting the other states. This
may be realised by a \emph{wire grid polarizer}
(\cite{Hec75}:105-106): a set of closely spaced fine conducting
wires. If a wave interacts with these wires, the electric field
component parallel to the wires drives electrons along the wire,
generating an alternating current which encounters resistance;
this absorbs energy from this component of the incoming field,
heating the material; the electrons re-radiate a wave which
further tends to cancel this component of the incident wave, while
the transverse component is not so affected. Hence the transmitted
wave is linearly polarized. The same effect occurs in a
\textit{polaroid polarizer}, consisting of many parallely aligned
microscopic crystals embedded in a transparent polymer film
(\cite{LipLip69}:132-133;\cite{Hec75}:105). Similarly a
spin-polarized current in a metal can be generated by passing the
current
through a ferromagnetic material.\\

A different example is a \emph{filter} that absorbs some wavelengths of light and transmits others, because of the
molecular structure of the glass, hence selecting a particular frequency range by adaptive absorbtion. \\

\begin{center}
\begin{tabular}{|c|c|c|}\hline
     \emph{\textbf{Classical Apparatus}} &Non-linear system & Non-unitary\\ \hline
    \emph{ Emergence} $\Uparrow $ &   \emph{Contextual effects} $\Downarrow$&  \emph{\textbf{Adaptive selection}}  \\
  \hline    \emph{\textbf{Quantum systems}}& State vector selection &  Non-unitary\\ \hline
\end{tabular}\\
\end{center}

{}\\ \textbf{Diagram 9:} \emph{The postulated contextual view of
state vector
preparation.}\\

\textbf{Emergence and top-down action}: In each case, the underlying
unitary quantum electrodynamics leads to emergence of higher level
classical structures (wires, crystals, and so on) that can then act
down to the particle level to cause non-unitary transformations
which can change a mixed incoming beam to a pure state (Diagram 9). As in
the case of the band structures of metals, this top-down action
depends on the physical structure of the polarizing material or
device as indicated in the above examples, and so is a case of
top-down causation by adaptive selection in the context of the
structure of the material. In the case of separation and selection,
the lower level evolution is unitary until selection takes place. In
the case of selective absorbtion, the ongoing non-unitary nature of
the resulting higher level effective action is reflected in an
energy loss and heating associated with the process.

\subsubsection{Measurement}\label{meas_td}
Measurement is a process with significant parallels to the process of state preparation, as just
pointed out. The experimental viewpoint is that the macro observer and apparatus have an existence
as macro entities that can be taken for granted, and that can influence states both in terms
of state preparation, and in terms of determining the outcomes of a measurement, for example
by choosing the axes along which spin will be measured. These are of course both cases of
top-down causation.\\

Does it go further than this: is the measurement process itself in
some sense also a case of top-down causation? In section
\ref{sec:measure context}, I will show that this is indeed so in
that the non-unitary measurement process is enabled by top-down
action from the structure of the detector to the particle
interactions. Here, I want to make just one other point: some of the
more advanced measurement techniques seem to directly involve
adaptive selection. For example this occurs in \emph{weak
measurements}, which are based in post-selection
(\cite{AhaRoh05}:225-227,230-235). This kind of selection of some
outcomes and discarding others is also central to the
\emph{generalized theory of quantum measurement} characterized by
Breuer and Petruccione (\cite{BrePet06}:83-85). It may well be worth
pursuing the idea that adaptive selection is the heart of the
measurement process (see Section 8).

\subsubsection{The arrow of time}\label{time2}
A further very significant case of top-down causation is the
determination of the arrow of time. It is a major topic, dealt with
in a companion paper \cite{Ell11a}. The picture that emerges from
the discussion there is shown in Diagram 10.\\

\begin{center}
\begin{tabular}{|c|c|c|}\hline
   \multicolumn{3}{|c|} {\emph{\textbf{The Arrow of Time}} }
   \\ \hline
   \textbf{Cosmology} &   &  \textbf{Brain, Society} \\ \hline
   \emph{Top-down effects }  $\Downarrow$ &      & $\Uparrow$ \emph{Bottom-up effects }   \\ \hline
   \textbf{Non-equilibrium environment} &  $\Rightarrow$  & \textbf{Molecular processes} \\ \hline
    \emph{Top-down effects } $\Downarrow$ & & $\Uparrow$ \emph{Bottom-up effects } \\ \hline
   \textbf{Quantum Theory} &  $\Rightarrow$  & \textbf{Quantum Theory} \\ \hline
\end{tabular}\\
\end{center}
 {}\\ \textbf{Diagram 10:} \emph{Contextual determination of the
arrow of time cascades down from cosmology to the underlying micro
processes, on the natural sciences side,
and then up to the brain and society, on the human sciences side}.\\

In summary: this view proposes that
\begin{itemize}
\item Spacetime is an evolving block
  universe, which grows as time evolves \cite{Ell06}. This fundamental arrow of
  time was set at the start of the universe.
  \item The observable part of the universe started off in a special state
  which allowed structure formation to take place and entropy to
  grow.
  \item The arrow of time cascades down from cosmology to the quantum
level (top down effects) and then cascades up in biological systems
(emergence effects).
\item There are an array of
technological and biological mechanisms that can detect the
direction of time, measure time at various levels of precision, and
record the passage of time in physically embodied memories.
  \item These are irreversible processes that occur at the classical level,
  even when they have a quantum origin such as a tunneling process, and
  at a foundational level must based either in a time-irreversible quantum
  measurement process or are a consequence of the special
initial state and the coupling of the atom to an infinite number of
electromagnetic degrees of freedom.
  \item  In conceptual terms they are the way the arrow of time parameter $t$ in the basic
equations of physics (the Dirac and Schr\"{o}edinger equations
(\ref{evolution}), Maxwell's equations and Einstein's equations on
the 1+3 covariant formulation \cite{EllMaaMac11}) is realised and
determines the rate of physical processes and hence the way time
emerges in relation to physical objects.
  \item Each of these processes is enabled by top-down action taking
  place in suitable emergent local structural contexts, provided by
  molecular or solid-state structures. These effects could not occur in
  a purely bottom-up way.
\end{itemize}
The detailed argument is in \cite{Ell11a}.

\section{The Measurement issue and contextuality}\label{sec:measure context}

Underlying the flow of time is the quantum measurement process. The
point to be made now is that a  measuring apparatus such as a Charge
Coupled Device (CCD) is a classical object. That is why it is able
to produce a specific measurement result --- it is not a quantum
system. How is this possible?  The resolution I propose is that a classical system emerges from the
underlying quantum components (see Section \ref{classmeas} below),
for example through the arbitrary allowed potential terms $V(x)$
(Sections \ref{sec:non-linear1} and \ref{effective}), and then acts
top-down on the quantum elements of the system to make a measurement
take place. Hence it is a contextual effect. The way this works is
set out in Diagram 11, with obvious similarities to Diagram 9.
Philosophically, the difference between state preparation and
measurement is that the outcome is largely determined by the
experimenter in the former case, but to a lesser
degree in the latter case.\\

\begin{center}
\begin{tabular}{|c|l|}\hline
     Non-linear system & \emph{\textbf{Classical Apparatus}} \\ \hline
    \emph{ Emergence} $\Uparrow $ &  $\Downarrow \emph{Contextual effects}$   \\
  \hline    Linear components & \emph{\textbf{Quantum systems}}   \\ \hline
\end{tabular}\\
\end{center}

{}\\ \textbf{Diagram 11:} \emph{The contextual view of quantum
measurement. Linearly acting quantum systems are assembled in a
non-linear way to create a classical apparatus with non-linear state
space, and non-Hamiltonian (non-unitary) evolution emergent from the
underlying physics (as discussed above). This macro apparatus acts
down on the micro quantum system being monitored by the
experimenter, resulting in both non-unitary state preparation, and a
set of specific measurement events where non-unitary state vector
projection takes
place. }\\

Section \ref{sec:reduction} considers the way state vector reduction
is related to context in general, and Section \ref{sec:photon}
fleshes this out in the case of photon detection.

\subsection{Contextuality and state vector reduction}\label{sec:reduction}
 Real experiments, such as the
Haroche single photon measurement (\cite{WisMil10}:45), involve
classical apparatus such as ionization detectors. These provide
the context within which measurements take place. Wiseman and
Milburn ask (\cite{WisMil10}:98)
\begin{quote}
``Should we include these as quantum systems in our description? No,
for two reasons. First, it is too hard. Quantum systems with many
degrees if freedom are generally intractable. ... Second, it is
unnecessary. Detectors are not arbitrary many body systems. They are
designed for a particular purpose: to be a detector. this means that
despite being coupled to a large environment, there are certain
properties of the detector that, if initially well defined, remain
well defined over time. These classical like properties are those
that are \emph{robust} in the face of decoherence... one of those
properties is precisely the one that becomes correlated with the
quantum system and so constitutes the \emph{measurement result} ''
\end{quote}

\noindent This emphasizes that the detection is a result of the
detector structure. Considering it as a classical system, the way
the measurement takes place depends on the physical details of this
detector, which is the local context for the measurement, for
example determining which spin component is measured. Thus this is
what one should concentrate on, to flesh out the abstract
concept of measurement embodied in the rule (\ref{trans}).\\

In what follows I will concentrate on photon detection, in order to
be definite. In this case, we have the following proposal:\\

\textbf{Thesis: The measurement process depends on the local context}. \\
\emph{Measurement (collapse to an eigenstate of some variables of
the system) occurs whenever the local context of the detector
structure causes such a projection to reliably take place in the
case that a photon impinges on an electron located in the detector.}
\footnote{Examples such as the quantum eraser and delayed choice
experiments show that the issue of ``when'' the detection takes  is
a subtle issue; c.f. \cite{EllRot10}.}\\

\noindent I explore this view, in accord with Landsman's review of the Bohr-Einstein debate \cite{Lan05},
 in some detail below. A similar discussion could be given for particle detection, magnetic
field detection, and so on.

\subsection{Photon detection}\label{sec:photon}

\textbf{What characterizes a measurement} (at the micro level)? When
does the interaction between a photon and an electron amount to a
measurement? When is it just scattering, and when is it absorption
of energy by the electron leading to the photoelectric effect as
part of a measurement process?
It may be either an active measurement
process or a passive measurement process 
depending on the context.

\subsubsection{Contexts}
A range of contexts is as follows:
\begin{itemize}
  \item \textbf{Plasma}: Electron in plasma: free electrons are not bound to nuclei, so interaction involves only
  an electron and a photon; Rayleigh scattering takes place (\cite{Dit58}:656-660), \cite{AloFin71}:14-20,\cite{ItzZub80}:224-230,286);
  \cite{PesSch95}:158-167). This heats up the plasma.
  \item \textbf{Gas}: Electron in free atom: (i) a photon does not change the state of the atom (Rayleigh
  and Compton scattering: \cite{Dit58}:656-659), (ii) changes the orbital level of the electron
  (\cite{Dir58}:175-178,239-248; \cite{CraThi84}:86-93), or (iii) frees it and so ionizes the
  atom (\cite{AloFin71}:30-31,\cite{CraThi84}:105-107) and thus ionizes the gas (\cite{Len86}:151-153). This
    leads to heating of the gas and reradiation of energy (\cite{CraThi84}:94-98).
  \item \textbf{Passive surface}: Electron in a physical structure where the
  photon is absorbed on interacting with the electron, but this does not free the electron.
  The surface heats up, which effect can be used
 to create a bolometer (\cite{Len86}:180-182, \cite{Chr10}:269-272), and re-radiates light,
  which makes it visible; this enables indirect measurement (\cite{BrePet06}:93).
  \item \textbf{Active surface}: Electron in a physical structure that
  absorbs a photon and is thereby freed from that structure (the photo electric effect), and then is used in a structure (a detector of some
  kind) to generate specific classical effects. This is the context in which photon \emph{detection} occurs, rather than just an interaction.
  \end{itemize}

\noindent Note
  that the kinds of calculation to determine the effect are quite different in the different cases listed here.
  It is the latter the constitutes an actual detection; only this case constitutes an active measurement.
  One can contrast this with
the way measurement is expressed in quantum theory texts in terms of operators and
  eigenvalues (cf. Section \ref{sec:basic}). That is the basis for what happens; this is where it becomes real.

\subsubsection{The Photo electric effect}
The basis of detection devices is the photoelectric effect
(\cite{AloFin71}:11-14), which occurs if an electron in a surface
absorbs the energy of a photon and thus has more energy than the 
work function  (the electron binding energy) of the material. It is
then ejected and produces a freely moving electron; if the photon
energy is too low, the electron is unable to escape the material
(\cite{AloFin71}:526; \cite{Zim79}:336-343; \cite{Chr10}:227-229;
\cite{Len86}:148-151).

\begin{quote}
\textbf{Detection} \emph{is when a photon impacts a 
structure
and causes an electron to be released which then causes a specific
physical effect on the structure. It is non-linear because there is
a detection threshold below which
no signal is detected.}
\end{quote}

All of this is a statement that what happens depends on the local
context: the "work function" is a macro property, depending on the
nature of the material (\cite{Zim79}:196-199). In the case of a
metal, the periodic crystal structure leads (via Bloch's theorem,
(\cite{Zim79}:16-20)) to the electronic band structure depending on
the particular solid involved (\cite{Zim79}:93-94,119-128). That is
the origin of the work function associated with the particular
metallic structure and associated optical properties
(\cite{Zim79}:255-291). The specific outcome is a result of the
layered atomic structure in which the electron is imbedded, which
creates the electronic band structure and work functions. Unlike the
case of free electrons, because these conduction electrons are in
the context of a crystalline structure, energy and momentum are not
conserved for the electron-photon pair; this is because the crystal
absorbs energy and momentum (\cite{Zim79}:60-61). This is at the
heart of why these processes are not unitary. As in Section
\ref{class2}, an open system can evolve
non-unitarily and with loss of energy since energy goes into environmental degrees of freedom.\\

Increasing the intensity of the light beam increases the number of
photons in the light beam, and thus increases the number of
electrons excited, but does not increase the energy that each
electron possesses. The output does not depend linearly on the
input: it has discrete steps in it because nothing is emitted up to
threshold intensity. Hence there is no superposition or entanglement
(Section \ref{class3}). The equivalence classes characterizing this
as top-down action are a consequence of Bloch's
theorem (see the remark on equivalence following (1.41) in \cite{Zim79}).\\

There
are many calculations of how photo-ionization arises from QED, e.g.
(\cite{Sch68}:420-422; \cite{Lou83}:179-184; \cite{Sak94}:339-341), but very few looking
at the photoelectric effect when the electron is in the band
structure in a solid (e.g. \cite{Zen08}). And these are statistical
calculations- they do not show how the wave function collapses in a specific interaction event.

\subsubsection{Types of Detectors}\label{sec:detectors}
The different types of photon detector
include the following; as indicated, each arises out of well understood quantum processes.
\begin{enumerate}
  \item \textbf{Photographic emulsions} Photographic plates (\cite{Len86}:175-177) record images
 via chemical reactions induced in the photographic emulsion
 by the photochemical effect (\cite{Len86}:150-151). Grains of silver
 bromide ($Ag^+ Br$) are imbedded in a transparent gelatin matrix; photons interact with
the grains to turn them into silver. When radiation of the right
wavelength impacts a silver bromide crystal, a series of reactions
produce a small amount of free silver in the grain \cite{Mye10}.

Initially, a free bromine atom is produced when the bromide ion
absorbs a photon:
\begin{equation}\label{photo1}
Ag^+ Br  + h\nu  \rightarrow Ag^+ + Br + e^-
\end{equation}
The silver ion can then combine with the electron to produce a
silver atom.
\begin{equation}\label{photo2}
Ag^+ + e^- \rightarrow Ag^0
\end{equation}
The detection event is the splitting up of the bromide ion, so releasing a free electron.

 \item \textbf{Photon counters and Photomultipliers}
 A \emph{photon counter} contains a fine wire in a positively charged cylinder (\cite{Dit58}:555). A photon ejects an electron from the wire by the photoelectric effect, which generates a small pulse of current.
A \emph{photomultiplier tube} (PMT) is a vacuum device where a photocathode is held at
a large negative voltage (\cite{Len86}:161-162; \cite{Chr10}:260-262). When a photon hits the photocathode and
ejects an electron into the vacuum due to the photoelectric effect, the electron is accelerated
to a more positively charged electrode called a \emph{dynode}, coated with a material such as
\textbf{$CsKSb$} or \textbf{$BeO$} that easily releases
several electrons to the vacuum when hit by an single energetic electron (this is the electronic
variant of the photoelectric effect). A greatly multiplying cascade of electrons proceeds
down a chain of eight such dynodes and leads to an electric pulse at the anode of the PMT.
  \item \textbf{Charge-Coupled Devices (CCDs)}
A Metal-Oxide-Semiconductor (MOS) capacitor (\cite{Chr10}:219-221)
is a sandwich of a grounded block of p-type semiconductor, a thin
insulator layer of \textbf{$SiO_2$}, and a thin layer of metal held
at a positive voltage. It has an electronic band structure such that
when an electron-hole pair is created by a photon in the depletion
region in the semiconductor adjacent to the insulator,
photoelectrons are stored in a potential well. A CCD
(\cite{Len86}:171-173, \cite{Cre09}:351-355;
\cite{Chr10}:243-260,317-321) contains a two-dimensional array of
MOS capacitors (one capacitor per pixel) so that when an image is
projected onto it, each capacitor accumulates an electric charge
proportional to the light intensity at that location. After such an
exposure, electronic control circuits read out each pixel
successively to produce a sequence of bits in the output line.

A newer development is CMOS imagers (\cite{Zim79}:355-357) where
charge to voltage conversion takes place in each pixel.
  \item \textbf{Photodiodes} (\cite{Len86}:150,154-156; \cite{Zim79}:336-343;
  \cite{Leo10}:107; \cite{Chr10}:223-227)
A photodiode is a $p-n$ junction with a potential across it. When a
photon of sufficient energy strikes an electron in the diode, via
the photoelectric effect it creates
 a free electron and a positively charged hole in the region
 between the p-doped and n-doped layers. This generates a photocurrent which is the sum of the
dark current (without light) and the light current (\cite{van07}:Ch4.6)
  \item \textbf{Super-conducting tunnel junctions (STJ)} These tunnel effect junctions (\cite{Len86}:156)
  are the most sensitive light detecting diodes.
An STJ is a Josephson junction (two pieces of superconducting
material separated by a very thin insulating layer) with a bias
voltage applied to the superconductors and a magnetic field
applied parallel to the junction (\cite{Chr10}:229-232).  The
current caused by quasiparticles tunnelling across the barrier is
suppressed for voltages less than twice the superconducting energy
gap.  A single photon can break apart multiple Cooper pairs,
promoting electrons into excited states. These tunnel across the
insulator and produce a current pulse.
  \item \textbf{Plant Leafs} Photosynthesis (\cite{AloFin71}:29-30) occurs when a photon causes a transition of a chlorophyll
  molecule, situated in a light harvesting complex in a leaf, from its ground state to an excited state
  (\cite{CamRee05}:182-195).
  After a chain of energy transfers, an electron is transferred from a special $\alpha$-chlorophyll molecule
  to a primary electron receptor where it causes a redox reaction, which then sets up an electron transfer chain
  that releases NADPH and ATP to a Calvin cycle. Immediate loss of energy by fluorescence of the excited molecules
  is prevented because of their context: ``each photosystem - a
  reaction centre surrounded by light harvesting complexes -
  functions  in the chloroplast as a unit'' (\cite{CamRee05}:189). An isolated chlorophyll molecule
  simply re-radiates the energy as the
  photo-excited electrons drop back to their ground state.
  \item \textbf{Animal eyes} Photoreceptors in the eye harvest energy by phototransduction
  enabled by rhodopsin (\cite{RhoPfl96}:269-274, \cite{KanSchJes00}:508-522).
   The primary step
  in the process is photon absorbtion followed by isomerization in a
  $\pi$ to $\pi^*$ or $n$ to $\pi^*$ orbital transition occurring in the light absorbing portion
  of rhodopsin (\cite{Atk94}:597), changing 11-\emph{cis} retinal to All-\emph{trans} retinal.
  This is enabled by an 11-\emph{cis} \textbf{C=C} conjugate double bond, and proceeds by causing
  a conformational change in the opsin portion of rhodopsin, which triggers the further steps
  in the process (\cite{KanSchJes00}:511): the rhodopsin molecule activates further
  molecules that open sodium channels in a rod cell and so producing
  hyperpolarization of the cell, eventually transduced into action
  potentials that travel to the optic nerve.
\end{enumerate}
In the latter two cases it is molecules imbedded in biological
structures that act as detectors. These are obviously highly
non-linear structures, physically of a scale much larger than that
of the electron. They form the classical context for the electron
that turns the electron-photon interaction into a detection. Note
that major further issues arise as to how detectors are configured
(in photomultipliers, bolometers, spectrographs, interferometers for
example) to obtain specific information \cite{Len86,Chr10}, and how
the data obtained is then processed. This all happens on the
classical side of the classical-quantum cut, and so is not the
concern here.

\subsubsection{The non-linear nature of physical measurement processes}
What is clear is that \emph{none of the detection processes
considered here obey the linearity conditions essential for quantum
theory superposition to apply (see Section \ref{sec:linear}), even
though they are enabled through well understood underlying quantum
interactions.} As in the case of state preparation (Section
\ref{prep}), superposition does not take place in the state space
(that is after all the nature of the measurement process) due to the
dynamics induced by top-down effects caused by the local environment
provided by the structure of the detector. Hence the reason these
processes can be regarded as classical processes is that, because of
the way the context shapes the outcomes,
they don't satisfy the requirements of being unitary. \\

At a certain level, that is a resolution of
the measurement paradox (see Section \ref{sec:paradox}): there
simply is no reason to believe that quantum theory will apply to any
realistically represented measurement apparatus. The measurement problem arises
when the abstraction of the measurement process (Section \ref{sec:basic})
is separated from the reality of detection events as outlined here. When discussions
of measurement do become more realistic (e.g. \cite{WisMil10}:42-49), they usually do so by
implicitly invoking the Heisenberg cut: macro apparatus such as detectors are
present (e.g. \cite{Len86}:Fig 1.3) as sites where the actual measurement takes place,
via the kind of processes outlined here. Detection processes like those discussed above take
place because the structure of the detector
is designed to behave in a non-linear way. That is what enables
the non-unitary measurement.\\

This does not of course solve the issue of what if anything determines the specific outcome of that process:
it is agnostic re the source of quantum uncertainty.
But the discussion here, in conjunction with the examples in Section \ref{sec:not linear}, does indicate how
non-linear detection events can arise
from the underlying linear quantum processes.

\section{Emergence of classical systems}\label{sec:classical emerge}
One of the puzzling issues in quantum theory is how to make a
classical apparatus emerge out of quantum foundations. How large a
system can be described by quantum theory? Where does the
micro-macro cut take place? This is the inverse to the issue of
making as large as possible a system behave quantum mechanically: an
answer to the one implies the answer to the other.

Section \ref{sec:classical criteria} considers basic criteria for
when we may expect a classical system to emerge, and Section
\ref{cut} how this may relate to the classical-quantum cut. Section
\ref{sec:class applications} gives some applications of criteria
developed there to some contentious examples.

\subsection{The basic criterion}\label{sec:classical criteria}
We have seen that to create a higher level quantum system, we don't
only have to protect it from decoherence - we also have to isolate a
linear system from all the messy non-linear entities in the world
around. This ensures a context where linearity holds for this part
of the whole, so that the quantum nature of the components
comprising the system results in a quantum  nature of the system
itself when we coarse grain from smaller to system scales (Section
\ref{sec:linear emerge}). To get the possibility of a quantum
system, we need to create conditions
where the linearity conditions L1-L3 of Section \ref{emergeq} hold at the system level,
allowing both a linear state space and linear dynamics.\\

Conversely, if we want classical systems to emerge, we must create
conditions where these conditions are not fulfilled. Ways of doing
so were indicated in Section \ref{sec:not linear}, with specific
examples given in Section \ref{sec:detectors}. In particular, we can
note the following:\\

\textbf{Quantum Limits}:\label{limits} \emph{Purely quantum
behaviour will generically not be possible at any level of
description of an isolated system where there are equilibrium
states, dissipative effects, threshold effects, feedback loops
occur, or where adaptive selection takes place.}\footnote{An
experimenter can `reach down' and elicit quantum behaviour,
using cleverly designed apparatus: but this is a highly exceptional situation.}\\

We can therefore arrange for classical behavior to emerge by setting
a context
where one or other of these elements occurs. 
 Generically this will happen as we consider larger and larger systems,
which is one reason why it is so difficult to
make macroscopic quantum systems. Considering the above examples gives guidance as to when this will occur.

\subsection{The classical to quantum cut} \label{cut}

\subsubsection{Quantum effects}\label{queff}
On the basis of the above examples, one may suggest the following:\\

\textbf{Quantum dynamical effects} \emph{will mostly occur at the
molecular level; however it can with great care be extended to much
larger systems (maybe 100 km) by creating appropriately
linear systems, but this will not occur naturally}.\\

That the molecular level can  be reached is shown both by
investigations showing that quantum effects can occur in fullerenes
and biomolecules \cite{Breetal02,Hacetal03} and occur in radical-ion
pair reactions \cite{Kom10}. This is of course compatible with the
usual understandings of QM as being a theory normally applicable on
small scales as indicated by the de Broglie wavelength $\lambda =
h/p = h/(m_0v)$, which for thermalized electrons in a non-metal at
room temperature is about $8 \times 10^{-9}m$, while the smallest
molecules have a length of about $ 10^{-10} m$.
But note the important distinction:\\

\textbf{Applicability of quantum theory versus significance of
entanglement effects}: \emph{There are separate issues as to whether
entanglement effects (i) can exist, 
and (ii) are significant. The latter depends
on how large physical objects are relative to scales set by the
Planck constant $\hbar$. The former is a qualitative issue related
to the possibility of describing causality at a particular level in
the hierarchy (Table 1) by Hamiltonian dynamics. There is no chance
of entanglement
effects being significant if they can't exist due to one or other of the situations mentioned above}.\\

Thus relating scales to the Planck constant is important as far as
significance of quantum effects is concerned, but is not the whole
story.

\subsubsection{Exceptional cases?}\label{queff1}
There are a series of exceptions where quantum effects are
significant on larger scales than the molecular scales.\\

 \textbf{Entangled photons}
From the viewpoint put here, an essential part of the wonderful
experimental work establishing entanglement over distances of many
kilometers (e..g. \cite{Viletal08,Schetal09}) is the careful
construction of linearly interacting systems over these macroscopic
scales: this is the endeavor to extend the linear aspects of physics
emphasized in Section \ref{sec:linear} to these distances (for
otherwise entanglement on such distances would be impossible). This is possible
in these cases because photons are able to travel macroscopic distances in transparent media
with virtually no interaction. These are truly macroscopic versions of essentially quantum phenomena.\\

 \textbf{Interferometric quantum non-demolition experiments}
Each LIGO gravitational wave observatory is based in a L-shaped
ultra high vacuum system, measuring 4 kilometers on each side,
forming  a power-recycled Michelson interferometer with
Fabry--P\'{e}rot arms. Squeezed optical states are fed in and read
out by quantum non-demolition technology
\cite{BuoChe01a,BuoChe01,Kimetal02}. Hence this corresponds to
setting up quantum states on a scale of 4 km. This is possible under
similar conditions to the previous case: it is a quantum photon
state, enabled by ultra-high vacuum and rigorous filtering of
background noise. This kind of detector centres on a remarkable
creation of macro-scale quantum states under very artificial
conditions that enable linearity to hold on these scales \cite{KhaMiaChe09}.\\

 \textbf{Superconductors} Similar
comments regarding linearity apply, at much smaller scales,
regarding the drive to quantum computing and the search for high
temperature superconductivity: these also depend on isolating
linearly interacting degrees of freedom in a suitable system. One
might note here that ordinary (low temperature) superconductivity
cannot occur spontaneously in nature, because the universe is
permeated with black body photons whose present temperature is
2.75K, which sets a lower limit to the temperatures of naturally
occurring bodies; hence the low temperatures needed for
superconductivity cannot occur without human intervention.\\

However the issue now is, should we regard large superconducting
magnets such as at those at the Large Hadron Collider as single
multi-particle quantum systems, hence with one macro-scale wave
function describing their entire state, or rather as local small
scale quantum systems, acting together to give quantum-based
macroscopic behaviour? According to (\cite{Dur00}:105),
superconducting magnets on scales of meters are enabled by cooling
to a few degrees K and manufacturing imperfection free wires (in
accord with Section \ref{emergeq}). The bound Cooper pairs
resulting from individual electrons interacting with the crystal
lattice and the lattice interacting with the other electrons are
not localized at one place in space, but are represented by
wavefunctions within the metal that spread out over a range of as
much as $1\mu m$. which is more than 1000 times the distance
between the individual electrons in the superconductor. But this
is not a macroscopic scale interaction; hence superconducting
macro behaviour is obtained by a collection of many local
entangled wave functions rather than a macro-scale wave function.
The quantum classical cut in this case is at about
the $1\mu m$ level.\\

\textbf{Degeneracy pressure: White Dwarfs and Neutron stars}
\emph{White Dwarfs} are stars with masses about $1.2 M_\odot$, radii
between 3000 and 2000 km, and so densities of about $10^6$ gr/cc
$\simeq 1\, ton/cm^3$. They have stopped burning their nuclear fuels
and are supported almost entirely by the pressure of a degenerate
electron gas (\cite{Cha57}:412-451, \cite{ZelNov71}:271-279,
\cite{MisThoWhe73}:619,627). \emph{Neutron stars} have also stopped
burning their nuclear fuels, and are also of mass about $1 M_\odot$,
but with radii of about 10 km, so their densities are about
$10^{14}$ gm/cc. Their cores are almost pure neutrons, rather like
one nucleus of $10^{57}$ neutrons in a superfluid state , but with
enough protons to prevent decay and enough electrons to create
charge neutrality \cite{OppVol39}. They are supported against
gravity by pressure of degenerate neutrons (\cite{OppVol39}, \cite{ZelNov71}:279-285, \cite{MisThoWhe73}:619). \\

These stars are prevented from collapsing by electron and neutron
degeneracy pressure respectively, hence they are held apart by
pressure generated by Pauli exclusion principle. In broad terms,
possible quantum states, limited by exclusion principle because the
wave function is antisymmetric, fill up from the bottom to the Fermi
level due to exclusion principle. The Fermi-Dirac equation of state
results and degeneracy pressure acts to stabilize the star
(\cite{Cha57}:357-402);\cite{ZelNov71}:163-188). So the system is
demonstrating quantum state effects on scales of 10km to thousands
of km.\\

The same issue arises as for the superconducting magnets: is there
one antisymmetric wave function for the states of the star as a
whole, with its energy levels filling up to generate the needed
pressure, or are there effective local boxes where degeneracy
pressure is generated, the star as a whole being held up by the
combined degeneracy pressures generated in all the little boxes? In
this case there is no global wavefunction for the star: the
antisymmetric wave function is only locally applicable. Discussions
of these stars \cite{Cha57,OppVol39,ZelNov71,MisThoWhe73} are
ambiguous on this issue. Of course the real physics of degenerate
gases is very complex (\cite{FetWal03}: 21-31,120-170) with nuclear
matter (\cite{FetWal03}:341-388,503-577) a
model for the effects one might expect in neutron star cores.\\

The issue is what is the relevant antisymmetric quantum state to
which the Pauli exclusion principle can be applied
(\cite{Cha57}:382-384). It seems reasonable to assume one only needs
this antisymmetry of states for nearby electrons in a white dwarf:
swapping it with one far distant will be irrelevant to real physical
behaviour. That is, the asymmetry \be \label{skew} \Psi(q_1,q_2,q_3,
...., q_N)= - \Psi(q_2,q_1,q_3, ...., q_N)\ee need only apply when
the particles $q_1$, $q_2$ are neighbouring particles (it is true
that if (\ref{skew}) holds for all neighboring particles, it will
also hold for arbitrarily distant ones; but that will be a
physically irrelevant byproduct of the significance of physical
crucial interchange asymmetry of neighbouring particles). This
suggests that local skew quantum state functions will suffice to
derive local classical gas properties (\cite{Cha57}:360-362) that
then get combined to determine the overall star structure; there
need be no global wave function for the star as whole, even thought
the degeneracy pressure can be thought of as being based in filling
available electron states for the star as a whole. \\

This conclusion
is supported by the fact that the local gas properties vary across
the star, so can hardly all be described as in the same state, and
by the use of a modified form of the Bethe-Goldstone picture for two
interacting nucleons in a Fermi sea, providing a qualitative basis
for the independent particle model of nuclear matter
(\cite{FetWal03}:358-366) developing out of the Hartree-Fock
approximation (\cite{FetWal03}:121-127). Eddington has emphasized
beautifully \cite{Edd26} the hurly-burly nature of what goes on in a
stellar interior: hardly a benign place to maintain quantum
entanglement. Accordingly I suggest the\\

\textbf{Local Degeneracy Hypothesis}: \emph{the physics of macroscopic
objects held apart by degeneracy pressure is determined by local
skew-symmetric state functions in boxes of sufficient scale to
determine a hydrodynamic approximation, rather than a global wave
function for the degenerate core as a whole}.\\

How large an averaging box is needed? Andre Peshier points out that
in heavy ion collisions, a hydrodynamic or thermodynamic
approximation is used when one has as few as 100 interacting
entities. This might be a reasonable estimate also for the cases of
what is required for the averaging volumes in white dwarfs and
neutron stars. \\

\textbf{Black Holes and Inflation}: The same issue arises also as to
whether quantum mechanics can be applied to black holes of arbitrary
size (following Hawking \cite{Haw84}) or the early universe (as in
inflation \cite{Dod03}). My suggestion will be the same: local
quantum mechanical effects everywhere will give the desired
consequences, without requiring a global wave function that applies
everywhere (although this might happen: such a global wave function
might be an emergent property of the system as a whole). This is a
proposal that needs testing.\\


\textbf{Overall} Andrew Briggs comments (private communication),
``we have very little experience of large entangled systems, indeed
it is an open question whether there is an upper limit of
`macroscopicness' (whatever that might mean) for a system to exhibit
quantum superposition (and hence entanglement). We are a very long
way from this in the laboratory, priding ourselves (I speak of the
community as a whole) in creating entanglement between, say, eight
trapped ions.'' I suggest that the examination of possible exceptions in
this section supports the view in the previous section:\\

\textbf{The classical quantum cut}: \emph{With a few rare carefully
engineered exceptions (which cannot occur naturally), the classical
quantum cut is at the molecular level or below.} The exceptional
cases can extend quantum states up to the order of $10-10^2$ Km.

\subsection{Applications}\label{sec:class applications}
Immediate corollaries of this discussion and the examples in Section \ref{sec:not linear} are,
\begin{itemize}
  \item \textbf{Corollary 1}: generically, systems in thermodynamic equilibrium will not exhibit quantum behaviour at a macroscopic scale
  (because the effective laws describing their macroscopic behaviour are classical laws);
  \item \textbf{Corollary 2}: generically, systems with threshold effects will not exhibit quantum behaviour at a macroscopic scale
  (because superposition does not apply across the threshold);
  \item \textbf{Corollary 3}: generically, living cells will not exhibit quantum behaviour
(because there are thousands of feedback loops in a living cell);
  \item \textbf{Corollary 4}: generically, animal brains will not exhibit quantum
behaviour (because they are complex networks involving both feedback loops and adaptive selection).
\end{itemize}
All these (complex) systems will ``typically'' not exhibit quantum
phenomena. Given a clever experimental design by a quantum
physicist, on some appropriately short time-scale and some
appropriately chosen subsystem, perhaps quantum effects should
become visible -- even in a living cell. But this is a highly
exceptional situation. There are literally thousands of processes
going on in  living cell. Quantum processes underlie them, and for
example tunnelling may take place. Genuinely quantum phenomena
such as entanglement are exceptional cases;\footnote{I am taking
for granted the stability of matter and the periodic table, as
classical outcomes.}  almost without exception these processes are
described in purely classical terms \cite{CamRee05}. This has
implications for well known controversies.

\subsubsection{Classical Measuring apparatus}\label{classmeas}
A long standing question is how it can be that one can construct a classically behaving laboratory apparatus out
of elementary particles whose behaviour is quantum-based. The arguments presented here suggest an answer:\\

A measuring apparatus is made of metals and other materials that are
in equilibrium states, hence Corollary 1 protects them from quantum
effects. Photon detectors rely on the photoelectric or related
photon effects, which rely on thresholds and so Corollary 2 protects
them. \\

It is a moot question as to whether the experimenter should be
regarded as part of the apparatus or not; in any case Corollaries 3
and 4 will help here, ensuring that the observer too is a
classically behaving system. Once detectors exist as classical
objects, they can exert a top-down influence on the detection
processes (Section \ref{sec:measure context}).

\begin{quote}
\textbf{Conclusion}:\emph{The conditions highlighted in the Corollaries above, based in the linearity requirements
 for the validity of quantum theory (Section \ref{sec:linear}), are sufficient to explain why a
classical observing apparatus can emerge from its underlying quantum components}.
\end{quote}
\noindent Actually it goes much further than that: the conditions highlighted in Section \ref{limits} are sufficient
to establish the existence of the classical world in general as a generic macrophenomenon, except under very
unusual circumstances (like an experimental setup that can generate entangled particle pairs over Km distances).
Thus they underlie the feature (emphasized in section \ref{sec:context}) that
\begin{quote}
\textbf{Classical reality} \emph{We can regard each of the higher levels of the hierarchy of complexity as a classical
domain, emergent from the underlying quantum theory but existing in its own right, with occasional quantum intrusions}.
\end{quote}

\subsubsection{Schr\"{o}dinger's cat}\label{cat} In their discussion (\cite{AhaRoh05}:121-124)
of the Schr\"{o}dinger's cat paradox,  Aharonov and Rohrlich
include the following representation of a final entangled state of
the cat and its environment (\cite{AhaRoh05}:eqn.(9.8)):
\begin{eqnarray}
\ket{\Psi(T)} &=& \frac{1}{\sqrt{2}} \ket{undecayed}\otimes\ket{untriggered}\otimes\ket{unactivated}
\otimes\ket{unbroken}\otimes\ket{live}\nonumber\\
&+& \frac{1}{\sqrt{2}} \ket{decayed}\otimes\ket{triggered}\otimes\ket{activated}
\otimes\ket{broken}\otimes\ket{dead}
\end{eqnarray}
One can certainly challenge the last term in each product, if not the earlier ones,
by considering the examples in Section \ref{sec:not linear} and the Corollaries above. The cat
will not exhibit quantum behaviour both because it is made of living cells, and has a brain.
\begin{quote}
\textbf{Conclusion}: \emph{Schr\"{o}dinger's cat can't be in a superposition because a Hamiltonian description
allowing the necessary unitary evolution does not apply to complex objects such as a cat}.
\end{quote}
Schr\"{o}dinger's cat states can however be constructed in quantum optics contexts (\cite{Leo10}:77,105).

\section{A View of the Classical World}\label{sec:conclusion}
\subsection{A viewpoint}
In the \emph{New York Review of Books}, Freeman Dyson wrote \cite{Dys11}
\begin{quote}
\emph{Toward the end of Feynman's life, his conservative view of quantum
science became unfashionable. The fashionable theorists reject his
dualistic picture of nature, with the classical world and the
quantum world existing side by side. They believe that only the
quantum world is real, and the classical world must be explained as
some kind of illusion arising out of quantum processes. They
disagree about the way in which quantum laws should be interpreted.
Their basic problem is to explain how a world of quantum
probabilities can generate the illusions of classical certainty that
we experience in our daily lives. Their various interpretations of
quantum theory lead to competing philosophical speculations about
the role of the observer in the description of nature.
Feynman had no patience for such speculations. He said that nature
tells us that both the quantum world and the classical world exist
and are real. We do not understand precisely how they fit together.
According to Feynman, the road to understanding is not to argue
about philosophy but to continue exploring the facts of nature."
}\end{quote}
This paper supports such a view. The basic theme is that a genuinely complex system is made up of simple systems, each
of which in isolation obeys linearity, but when assembled together
in a causal network their combination does not, the elements being
combined thus precisely in order to allow non-linear interactions
such as positive and negative feedback and adaptive selection. This prevents superposition of states, and hence quantum
phenomena will not be expected to occur on macroscopic scales. Macro-scale entities will exist as entities with
causal powers in their own right, thus enabling top-down causation to take place as well as bottom up.
Hence all the levels of emergent reality should be treated on an equal ontological basis:  none is a privileged level of
existence, all are equally real (see Denis Nobel's article in \cite{EllNobOCo11}):
\begin{quote}
\textbf{HYPOTHESIS 1}: \emph{Macrophysics exists on an equal basis to the micro.
It is just as real and just as causally effective}.
\end{quote}
This view is implicit in all quantum mechanics studies
where macro-concepts like a `photon detector' are used, often without comment. They are part of the experimental
machinery that must be taken for granted in order that experimental physics can proceed.\\

As a consequence, emergence and contextuality should be seen as a
key feature of science. On the one hand, we need to take the bottom
up emergence of higher level properties seriously, as we consider
the degree to which quantum theory may be applicable to higher
levels of the hierarchy of complexity. Some approaches at use in
present in effect don't do so: they implicitly assume this process
will lead to quantum behaviour at higher levels, when that
assumption may not be
true. \\

On the other hand, top-down influences crucially affect quantum
level outcomes, as for example in the process of decoherence.
Because top down action takes place, a concept of non-quantum macro
systems is essential in formulating quantum theory. This is
essentially the Copenhagen interpretation of QM.
\begin{quote}
\textbf{HYPOTHESIS 2}: \emph{Contextuality is crucial: one should see quantum behaviour as the result of
an interaction of bottom up and top down effects}.
\end{quote}
In other words, complexity is the key criterion in the
classicalisation of the universe. Leggett states,
 \begin{quote}
 ``QM is a very `totalitarian' theory, and if it applies to individual atoms
 and electrons, then it should prima facie equally apply to the macroscopic objects
 made up by them,
 including any devices which we have
 set up as a measuring apparatus''  \cite{Leg08}.
 \end{quote}
 By contrast, this paper proposes that the appropriate dynamics at higher levels
 is determined by coarse graining of the dynamics at lower levels (see Diagrams 1 and 2).
 In that case,
  QM will only apply to higher levels in the hierarchy under very restricted circumstances.

\subsection{Questions}\label{quest}
\textbf{In summary}, This paper has provided a broad framework to look at some issues in the relation
of quantum theory to the macro world, based on the proposals given in Section \ref{main}, and summarized in
diagrams 1, 2, and 8. This view
respects the reality check provided in Section \ref{sec:basis}, but
obviously leaves many questions unanswered. Particular issues to explore include,
\begin{itemize}
  \item\textbf{ Almost linearity}: Section \ref{sec:linear} has emphasized the need for linearity in order that quantum
  physics is applicable, but has not considered \emph{how} linear a system has to be: \emph{when is `almost linearity'
   adequate}?
  This is a key question, relating to such issues as spatial and temporal coherence. It relates to considering quantum theory as a theory of
  perturbations: many systems can be regarded as linear if one restricts the space, time, and energy scales enough,
  so the issue will be for how long and over what scales almost-linearity will be at acceptable levels. This is where the
  uncertainty principle will enter, and relates to issues such as to what degree genuinely quantum properties
  occur in biology \cite{Llo11,Bal11}. Put another way, if  complexity is the key criterion
  in the classicalisation of the universe how is complexity to be quantified
  for this job? Can it be done in a way that avoids introducing additional
dimensional constants into physics?
  \item \textbf{Detection Processes}: It will be useful to develop detailed QED models of the kinds of detection
  processes discussed in Section \ref{sec:detectors}, keeping careful track of precisely where the projection
  process (\ref{trans}) occurs and what contextual features constrain how it happens. This might possibly
  provide a framework for explicit context dependent collapse models as an alternative to those by Ghirardi et al
\cite{Ghi07} and Penrose \cite{Pen89,Pen04}, based in a top-down
process of adaptive selection (Section \ref{adapt}) because adaptive
selection underlies the dual process of state vector preparation
(Section \ref{prep}). This might possibly induce an extra term in
the Schr\"{o}dinger equation, in a way similar to the way effective
potentials arise (Section \ref{effective}). This raises a further
issue: if a time scale for collapse is introduced into the theory,
representing a new fundamental parameter in physics, how is this
constant related to the measure of complexity of the higher level?
\item \textbf{State Vector Preparation:} Investigating that proposal will be assisted by developing detailed QED models
of the process of state vector preparation, which as just remarked is based in a top-down process of adaptive selection
(Section \ref{prep}).
This will clarify how it provides a well-founded route, based in established quantum physics, that can
lead from mixed to pure states
 and is able to produce an effective collapse of the wave function, because it can produce eigenstates. A key
 issue here is clarity on precisely how the idea of state vector reduction relates to particle creation and
 annihilation as represented in QED.
 \item \textbf{Coarse graining and detection} Related to this is the fact that any coarse-graining
 implicitly involves both a temporal and spatial scale, and it will often also involve an energy scale.
 Thus for example a density measurement for a gas will correspond to a specific averaging length scale;
 an image obtained by a detector will correspond to specific angular, exposure time, and energy scales.
 Hence measuring coarse grained entities involves convolution with a detection function or window
 function. The effect of such coarse graining on detected entities will affect what we can actually measure,
 and it selects the information we gather from all the other incoming stuff we don't want (another form
 of adaptive selection). Such filtering of what we detect is essentially the start of pattern recognition,
 indeed sophisticated filters can implement genuine pattern recognition, thereby collecting useful
 information. Exploring these effects in relation to issues of emergence and information may be useful.
\item \textbf{Quantum theory and the arrow of time} As part of this project, it should be that the  time-asymmetry of the
quantum measurement process emerges in a contextual way. There seem to be two parts to this.  (a) The first is that a detection
  process depends on setting the detector into a ground state before detection takes place (analogously to the way computer memories have
to be notionally cleared before a calculation can begin). This is an
asymmetric adaptive selection process, whereby any possible initial
state of the detector is reduced to a starting state, thereby
decreasing entropy. It will be implemented as part of the detector
design. (b) The asymmetry of the collapse process may rely on the
fact that the future does not yet exist in a EBU \cite{Ell06}, hence
we cannot have advanced Green functions contributing to a Feynman
propagator. There does not seem to be any other plausible way to
relate the global cosmological arrow of time to the local arrow of
time involved in collapse of the wave function. This needs to be
elucidated; a start is made in \cite{Ell11a}.
\item \textbf{Test of non-linearities} This implied coarse graining in any detection relates to a precision
 test of quantum mechanics proposed by Weinberg \cite{Wei89}, based on searching for the detuning of resonant
 transitions in ${}^9Be^+$ ions. Such an experiment in effect involves a window function of scale the
 size of the  ${}^9Be^+$ ion, which has a radius of 1.12A. Extending such tests to larger scales is
 obviously extremely difficult; but still one might have that as a goal.
 \item \textbf{Inequalities}: Like many other studies, Leggett's ``macroscopic realism'' condition
\cite{Leg08} is based on the view that
quantum theory applies at all levels (as in Diagram 3). The present paper suggests this cannot be taken for granted;
higher level quantum behaviour will not generally emerge from the combination of lower level quantum systems.
Hence the Leggett-Garg inequalities and their generalizations \cite{Leg08} may be a way to test the relative
viability of these two approaches. Developing such tests would clearly be useful.
\end{itemize}

\noindent Perhaps the most unexpected feature emerging from this analysis is the conclusion that
\emph{adaptive selection may
play a key role in quantum physics, as well as in biology}. This conclusion is foreshadowed by the way it may be
seen as playing a key role in environmental decoherence \cite{Zur03,Zur04}, which is central to the emergence
of classical states.\\

Does this view regard unitary quantum physics as an essentially
fundamental theory (no exceptions are allowed)?\footnote{I am
indebted to Guenter Mahler for these questions and the following
comment.} No, it subscribes to the ``Leggett program'', according to
which one should expect inherent (fundamental) limits to quantum
behavior of higher level (more complex) systems. Unitary quantum
physics is fundamental in that it applies to everything at a
foundational level in the hierarchy of complexity, except when state
vector reduction takes place in consequence of a process that is yet
to be determined. This does not mean it necessarily applies at
arbitrary higher levels. That depends on the emergence of higher
level behaviour, which may or may not obey unitary quantum precepts;
indeed it is clear it often does not do so (as illustrated above by
many examples). That emergence is due to the state vector collapse
process
at the lower levels (as emphasized by Leggett).\\

How can one distinguish a fundamental collapse event from a
``standard'' environment-induced one? On this view, they are all
environmentally induced. How can one ever distinguish a fundamental
non-Hamiltonian behavior from an effective non-Hamiltonian behavior
resulting from a reduced description of an underlying Hamiltonian
system? Such effective descriptions abound. On this view, those
reduced descriptions that result in a non-Hamiltonian behaviour do
so by implicitly assuming a lower level state reduction mechanism
(underlying events such as such as an Umklapp-process). All
``phenomena'' are contextual, in the sense that  what we see depends
on our resources. Now take the resources to specify the observer;
this observer cannot be "exorcised", he is needed to
condition and select the appropriate physical description.\\

The standard view is that any isolated system can in principle be
described by a single wave function, no matter how large they are.
Often the wave function can be written as a product state of
wavefunctions of individual systems. If that cannot be done then the
individual systems are entangled; but decoherence will rapidly
remove such entanglement in realworld situations. If it can, then
these are the local wave functions that underlie local physics. This
paper proposes that this view must be
treated with caution: one should check if such a wave function emerges from
the micro level.\\

\noindent \textbf{The ultimate take home message} is three fold:
\begin{itemize}
 \item \emph{Considering issues such as state preparation and the process
  of state vector reduction should be done taking realistic contexts into account:}
  on the one hand, the cosmic environment in which we live (cf. \cite{Ell11a});
  on the other, the complexities of life as we experience it in the everyday
  world. When we tackle such issues, the abstractions of our scientific models must be rich
  enough to take this complexity seriously.
  \item \emph{It is not OK to just assume that a Hamiltonian formulation
will apply to any old system, no matter how large}. You can assume
it will apply to the component parts at the quantum level; but when
these parts are assembled into a complex system, a Hamiltonian
description may or may not be valid as a description of the higher
level dynamics. You have to investigate whether this is so or not.
In many cases the answer will be that it is not applicable.
  \item \emph{The complexity we see arises from a combination of
  bottom-up emergence of higher levels of behaviour, combined with top-down
  influences that determine the actual outcomes in specific contexts}.
  How do you tell when it is bottom up causation alone? -- when
  lower level action by itself leads to well-determined higher level behaviour,
  as in the perfect gas laws and the black-body radiation formula.
  How do you tell when top-down causation makes a significant
  difference to outcomes? When higher level effects such as the band structure
  of metals is the main determinant of the specific lower level
  outcomes, as in the case of superconductivity and semiconductors: you cannot determine
  the outcome on the basis of the lower level properties alone
  \cite{Lau00}.
\end{itemize}
One of the most important examples of top-down causation is the
existence and direction of the arrow of time. This is a crucial
feature of the daily world, without which we would not be here. The
accompanying paper \cite{Ell11a} makes the case that this key issue
is best studied by looking in detail at how physical systems,
arising out of the underlying  unitary physics, detect the one-way
flow of time. \\

\noindent \textbf{Acknowledgement}: \\

I thank Anton Zeilinger, Andrew Briggs,  Jeff Murugan, David
Aschman, Raoul Viollier, Andre Peshier, and particularly Paul Davies
for useful discussions, Guenter Mahler for helpful comments, Per
Sundin for a correction to a previous version, and particularly
Jeremy Butterfield for detailed comments on a previous version of
the paper. I thank an anonymous referee for detailed comments
that have improved parts of this paper.\\

I thank Anton Zeilinger for hospitality at meetings at the
International Academy, Traunkirchen, that were very helpful in
developing these ideas. I thank the National Research Foundation (South Africa) and
the University of Cape Town for support.
\newpage

gfre::version 2012-04-06
\end{document}